\documentclass[conference]{IEEEtran}
\usepackage[utf8]{inputenc}
\usepackage[T1]{fontenc}
\usepackage[english]{babel}
\usepackage{booktabs}
\newcommand{\ra}[1]{\renewcommand{\arraystretch}{#1}}
\usepackage{listings}
\usepackage[
 lambda,
 operators,
 advantage,
 sets,
 adversary,
 landau,
 probability,
 notions,
 logic,
 ff,
 mm,
 primitives,
 events,
 complexity,
 asymptotics,
 keys,
]{cryptocode}
\usepackage{amsthm}
\usepackage{xspace}
\usepackage{layouts}
\usepackage[binary-units=true]{siunitx}
\usepackage{subfigure}

\usepackage{makecell}

\newtheorem{defn}{Definition}
\newtheorem{cnst}{Construction}
\newtheorem{thm}{Theorem}
\newtheorem{lem}{Lemma}
\renewcommand{\kgen}{\pcalgostyle{Gen}}
\newcommand{\der}{\pcalgostyle{Der}}
\newcommand{\Msp}{\mathcal{M}}
\newcommand{\Csp}{\mathcal{C}}
\newcommand{\Isp}{\mathcal{I}}
\newcommand{\Lsp}{\mathcal{L}}
\newcommand{\ceghaseek}{\mathbb{G}}
\newcommand{\ceghasesk}{\ensuremath{\langle \mathbb{G}, \allowbreak q, \allowbreak g, \allowbreak a, \allowbreak x, \allowbreak y, \allowbreak h, \allowbreak j, \allowbreak k, \allowbreak D \rangle}}
\newcommand{\ceghaseski}{\ensuremath{\langle d_1, \ldots, d_t, d \rangle}}
\newcommand{\ceghasect}{\ensuremath{\langle u_1, \allowbreak v_1, \allowbreak \ldots, \allowbreak u_t, \allowbreak v_t, \allowbreak s, \allowbreak w \rangle}}
\newcommand{\elghaseek}{\mathbb{G}}
\newcommand{\elghasesk}{\ensuremath{\langle \mathbb{G}, \allowbreak q, \allowbreak g, \allowbreak a, \allowbreak x, \allowbreak y, \allowbreak h, \allowbreak j, \allowbreak k \rangle}}
\newcommand{\ufcpa}{\pcnotionstyle{UF\pcmathhyphen{}CPA}}

\renewcommand{\state}{\pckeystyle{st}}

\newcommand{\eg}{e.g.,\ }
\newcommand{\ie}{i.e.,\ }
\newcommand{\dfauth}{DFAuth\xspace}

\makeatletter
\newcommand{\mypm}{\mathbin{\mathpalette\@mypm\relax}}
\newcommand{\@mypm}[2]{\ooalign{%
  \raisebox{.1\height}{$#1+$}\cr
  \smash{\raisebox{-.6\height}{$#1-$}}\cr}}
\makeatother

\author{
\IEEEauthorblockN{Andreas Fischer}
\IEEEauthorblockA{SAP SE \\ Karlsruhe, Germany \\ andreas.fischer02@sap.com}
\and
\IEEEauthorblockN{Benny Fuhry}
\IEEEauthorblockA{SAP SE \\ Karlsruhe, Germany \\ benny.fuhry@sap.com}
\and
\IEEEauthorblockN{Florian Kerschbaum}
\IEEEauthorblockA{School of Computer Science \\ University of Waterloo, Canada \\ fkerschbaum@uwaterloo.ca}
\and
\IEEEauthorblockN{Eric Bodden}
\IEEEauthorblockA{Heinz Nixdorf Institute \\ University of Paderborn, Germany \\ eric.bodden@upb.de}
}
\title{Computation on Encrypted Data using Data Flow Authentication}

\begin{document}
\maketitle

\begin{abstract}
Encrypting data before sending it to the cloud protects it against hackers and malicious insiders, but requires the cloud to compute on encrypted data.
Trusted (hardware) modules, \eg secure enclaves like Intel's SGX, can very efficiently run entire programs in encrypted memory shielding it from the administrator's view.
However, it already has been demonstrated that software vulnerabilities give an attacker ample opportunity to insert arbitrary code into the program.
This code can then modify the data flow of the program and leak any secret in the program to an observer in the cloud via SGX side-channels.
Since any larger program is rife with software vulnerabilities, it is not a good idea to outsource entire programs to an SGX enclave.

A secure alternative with a small trusted code base would be fully homomorphic encryption (FHE) -- the holy grail of encrypted computation.
However, due to its high computational complexity it is unlikely to be adopted in the near future.
As a result researchers have made several proposals for transforming programs to perform encrypted computations on less powerful encryption schemes.
Yet, current approaches fail on programs that make control-flow decisions based on encrypted data.
In this paper, we introduce the concept of {\em data flow authentication} (\dfauth).
\dfauth prevents an adversary from arbitrarily deviating from the data flow of a program.
Hence, an attacker cannot perform an attack as outlined before on SGX.
This enables that all programs, even those including operations on control-flow decision variables, can be computed on encrypted data.

We implemented \dfauth using a novel authenticated homomorphic encryption scheme, a Java bytecode-to-bytecode compiler producing fully executable programs, and SGX enclaves.
We applied \dfauth to an existing neural network that performs machine learning on sensitive medical data. 
The transformation yields a neural network with encrypted weights, which can be evaluated on encrypted inputs in $0.86$ seconds.
\end{abstract}

\section{Introduction}
Many critical computations are being outsourced to the cloud.
However, attackers might gain control of the cloud servers and steal the data they hold.
End-to-end encryption is a viable security countermeasure, but requires the cloud to compute on encrypted data.

Trusted (hardware) modules, \eg secure enclaves such as Intel's SGX \cite{Intel_SGX3, Intel_SGX2, Intel_SGX1}, promise to solve this problem.
An SGX enclave can very efficiently run an entire program in encrypted memory shielding it from the administrator's view.
However, it already has been demonstrated that software vulnerabilities give an attacker ample opportunity to insert arbitrary code into the program \cite{MS_SGX_Attack}.
These attacks are called return-oriented programming (ROP) and piece together programs from code snippets before return statements in the actual program.
Since any larger program is rife with software vulnerabilities, it is hence not a good idea to outsource entire programs to an SGX enclave.
The inserted code can modify the control and data flow of the program and leak any secret in the program to an observer in the cloud via SGX side-channels \cite{206170,203698,Schwarz2017}.

In particular, consider the following data flow modification attack that efficiently leaks a secret $x$ in its entirety.
We assume an encrypted variable $\enc(x)$ in the domain $[0, N-1]$ is compared to $N/2 - 1$.
The ``then'' branch is taken if it is lower or equal; the ``else'' branch otherwise.
This can be observed, for example, by the branch shadowing attack presented in \cite{203698}.
The observation of this behaviour leaks whether $x \leq N/2 -1$.
This becomes quite problematic when assuming a strong, active adversary that can modify the control and data flow.
The adversary may then create constants $\enc(\bar{x})$ for $\bar{x} \in \{ N/4, \allowbreak N/8, \allowbreak N/16, \allowbreak \ldots, \allowbreak 1 \}$ in the encrypted code, add those to the variable $\enc(x)$ and re-run the control-flow branch.
This way, by consecutively adding or subtracting the constants, the adversary can conduct a binary search for the encrypted value.

As a defence for this attack (of modifying the data flow), we introduce the concept of {\em data flow authentication (\dfauth)}.
We equip each control-flow decision variable with a label (loosely speaking: a message authentication code), such that only variables with a pre-approved data flow can be used in the decision.
Variables carry unique identifiers that are preserved and checked during the encrypted operations.
This prevents an adversary from deviating from the data flow in ways that would allow attacks such as the one we mentioned before.
Note that a program may still have {\em intentional} leaks introduced by the programmer.
However, \dfauth restricts the leakage of any program to these intended leaks by the programmer which the programmer could avoid by using appropriate algorithms such as data-oblivious ones.
In essence, the technique restricts the information flows to those that are equivalent to the original program's information flows.
We give a definition in Section~\ref{sec:coed}.

Fully homomorphic encryption (FHE) \cite{Gen09} would be another alternative to compute on encrypted data without leaks.
Due to its disappointing performance \cite{GenHal12}, researchers are seeking efficient alternatives that offer similar security.
Fortunately, we know how to efficiently perform additively and multiplicatively homomorphic operations on encrypted data.
Furthermore, if we reveal the control flow of a program (instead of computing a circuit), efficient computation seems feasible.
Note that any control flow decision on an encrypted variable is an intentional leak by the programmer.
Several proposals for program transformation into such encrypted computations have been made.
MrCrypt \cite{TetLes13}, JCrypt \cite{DonMil16} and AutoCrypt \cite{ACCCS13} each offer an increasing set of programs that can be computed on encrypted data.
To support encrypted computation on all programs, however, one needs to convert between different homomorphic encryption schemes.
These conversions are very small routines, such that we can scrutinize their code and implement them safely in a \emph{Trusted Module} likely without any software vulnerabilities.

In this way we combine the benefits of partially homomorphic encryption with a small code base and the efficiency of unprotected program execution.
Our re-encryption modules are small and program-independent and are run protected in the SGX enclave whereas the program runs efficiently on homomorphic encrypted values in unprotected memory.
We take care not to destroy the benefits of outsourcing.
The verification of labels is constant time and does not dependent on the homomorphic computation.
To this end we introduce our own authenticated homomorphic encryption scheme HASE.

We implemented the program transformation in a bytecode-to-bytecode compiler, such that the resulting programs are executable.
We evaluate \dfauth based on two applications: a checkout (shopping cart) component of a sales application and a neural network performing evaluations on sensitive medical data.
The transformed applications execute in \SI{2.3}{\milli\second} and \SI{0.86}{\second}, respectively.
This shows that \dfauth is practically deployable, although it provides extensive security guarantees.

In summary, our contributions are:
\begin{itemize}

\item We define the concept of {\em data flow authentication} and show its interference equivalence property in a program dependency graph.

\item We present a new {\em authenticated homomorphic encryption scheme} HASE, which can be used for constant time implementation of data flow authentication.

\item We implemented and evaluated a {\em bytecode-to-bytecode program transformation} for computation on encrypted data using data flow authentication.

\item We implemented and evaluated transformed programs, \eg machine learning, using {\em Intel's SGX}.

\end{itemize}

This paper is structured as follows.
In the next section, we provide our adversary model and various definitions of our authenticated homomorphic encryption scheme HASE required in subsequent sections.
In Section~\ref{sec:coed}, we introduce data flow authentication \dfauth and the security it provides.
Section~\ref{sec:hase} presents our HASE constructions and discusses their security.
Details about our implementation in Java are given in Section~\ref{sec:impl} and Section~\ref{sec:eval} shows the results of our evaluation using this implementation.
Section~\ref{sec:rel} presents related work before Section~\ref{sec:con} concludes our work.

\section{Definitions}
\label{sec:prel}
In order to understand the security of data flow authentication, we first define the overall adversary model we consider, the algorithms that HASE offers and the security it guarantees.
%Section~\ref{sec:prel:not} presents notational subtleties, Section~\ref{sec:prel:games} recalls game-based security and Section~\ref{sec:prel:ass} provides security assumptions used in subsequent sections.

\subsection{Adversary Model}
\label{sec:adversary}

\begin{figure}[bt]
\centering
\includegraphics[width=\columnwidth]{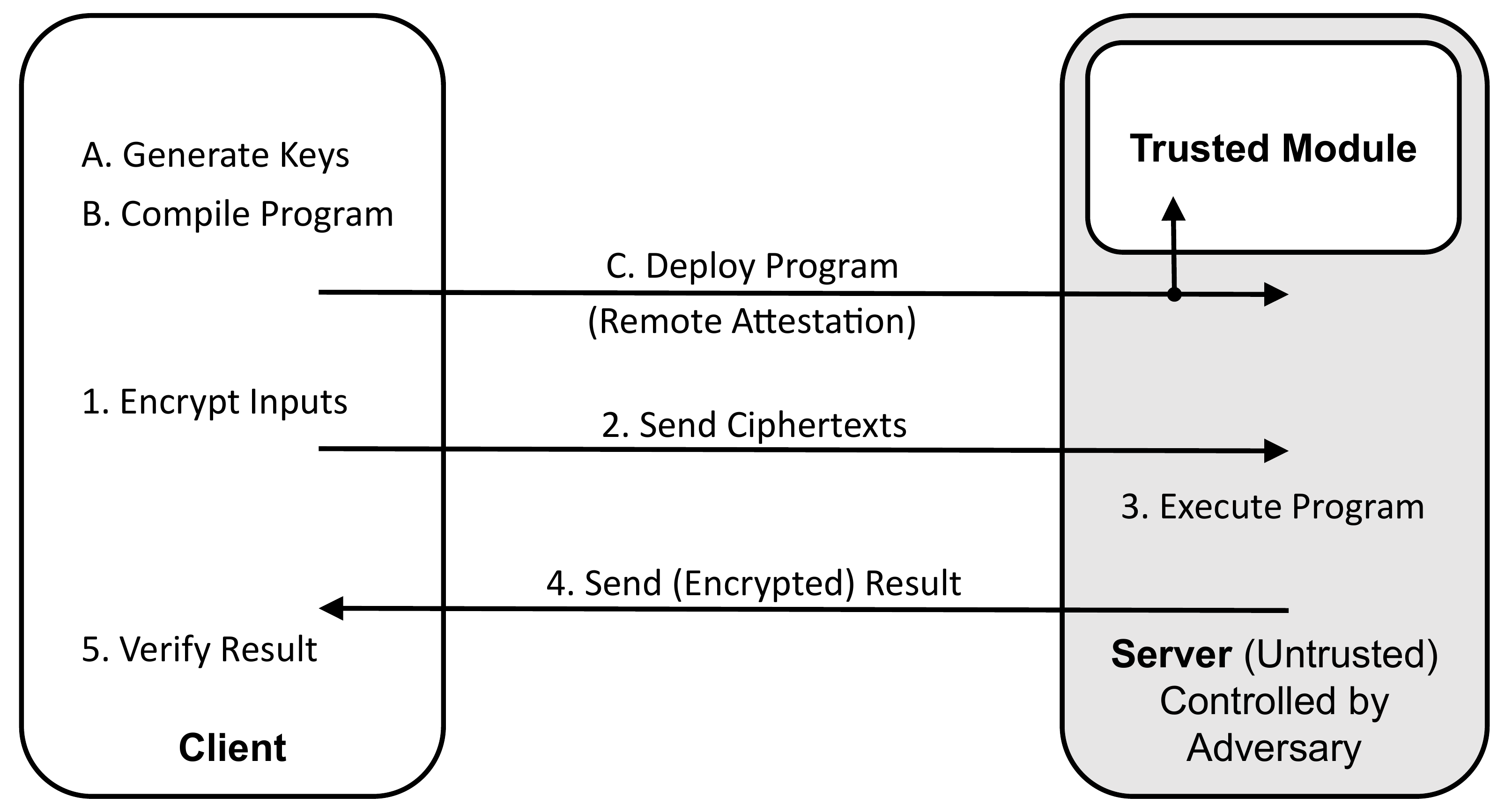}
\caption{System Overview}
\label{fig:overview}
\end{figure}

We consider a scenario between a trusted client and an untrusted cloud server, which has a Trusted Module, \eg an Intel SGX enclave.
Figure~\ref{fig:overview} depicts the process and its trust boundaries.
The client wishes to execute a program at the cloud server with sensitive input data.
Our security objective is to leak only the information about the inputs to the cloud server that can be inferred from the program's executed control flow.

We distinguish two phases of this outsourced computation: setup and runtime.
First, the client chooses the keys for the encryption of its inputs in our HASE encryption scheme (A).
Then the client transforms the intended program using a specialized HASE-enabled compiler and uploads it to the cloud (B).
The server deploys some parts of the program into the Trusted Module which the client verifies by remote attestation (C).
This concludes the setup phase.

In the runtime phase, the client can execute -- multiple times if it wishes -- the program on inputs of its choice.
It (1) encrypts the inputs using the information from the compiled program and (2) sends the ciphertexts to the cloud server.
The cloud server now (3) executes the program.
We assume an active adversary controlling the cloud.
The adversary can
\begin{itemize}

\item {\em read} the contents of all variables and the program text (except in the Trusted Module).

\item {\em modify} the contents of all variables and the program (except in the Trusted Module).

\item {\em continuously observe and modify} the control flow, \eg by breaking the program, executing instructions step-by-step and modifying the instruction pointer (except in the Trusted Module).

\item do all of this arbitrarily {\em interleaved}.

\end{itemize}
After the execution of the program the server (4) returns an encrypted result to the client.
The client can then (5) verify the result of the computation.

We ensure the following security property:
The server has {\em learnt nothing beyond the intended information flow} of the program to unclassified memory locations (interference equivalence\footnote{See Section~\ref{sec:ie} for a formal definition.}).

\subsection{Notation}
\label{sec:prel:not}
% Objects and Method calls
To denote an object whose members can be unambiguously accessed individually we write $\Angle{\ldots}$.
We use $.$ to access object members, for example $O.A()$ refers to an invocation of algorithm $A$ on object $O$.

% Deterministic assignment, comparison, non-deterministic assignment,
% uniform sampling.
We use $:=$ for deterministic variable assignments and $=$ for comparisons.
To indicate that an output of some algorithm may not be deterministic we use $\gets$ instead of $:=$ in assignments.
We write $x \sample X$ to sample $x$ uniformly at random from a set $X$.

% [m, n] set notation
For $m, n \in \mathbb{N}, m < n$ we use $[m, n]$ to refer to the set of integers $\{m, \ldots, n\}$.
% https://en.wikipedia.org/wiki/Projection_(mathematics)
% https://en.wikipedia.org/wiki/Projection_(set_theory)
For a $k$-tuple $x = (x_1, x_2, \ldots, x_k)$ we refer to the projection of $x$ onto its $i$-th ($i \in [1, k]$) component as $\pi_i(x) := x_i$.
Similarly, for a set of $k$-tuples $S$ we define $\pi_i(S) := \{ \pi_i(x) : x \in S \}$.

% Groups
We follow the established convention of writing the group operation of an abstract group multiplicatively.
Consequently, exponentiation refers to a repetition of the group operation.
We may refer to a group $(\mathbb{G}, \cdot)$ simply as $\mathbb{G}$ if the group operation is clear from the context.

% Lambda, PPT, negligible function
Throughout the document $\secpar$ denotes a security parameter and $\secparam$ refers to the unary encoding of $\secpar$.
The abbreviation \emph{PPT} stands for \emph{probabilistic polynomial time}.
A function $f: \mathbb{N} \rightarrow \mathbb{R}^{+}$ is called \emph{negligible} in $n$ if for every positive polynomial $p$ there is an
 $n_0$ such that for all $n > n_0$ it holds that $f(n) < 1 / p(n)$.

% Oracle Functions
To indicate that some algorithm $\adv$ is given black-box access to some function $F$ we write $\adv^{F}$.
Each parameter to $F$ is either fixed to some variable or marked using $\cdot$ denoting that $\adv$ may freely choose this parameter.

\subsection{Game-Based Security}
\label{sec:prel:games}
% Game, Challenger, Adversary.
We provide security definitions as \emph{games} (security experiments) played between a PPT \emph{challenger} and a PPT \emph{adversary} $\adv$ \cite{Games}.
The result of the game is $1$ if $\adv$ wins the game (\ie breaks security) and $0$ otherwise.
% Advantage
$\adv$'s \emph{advantage} is defined as the probability of $\adv$ winning the game minus the probability of trivially winning the game (\eg by guessing
 blindly).
% Shorthand for success probability.
%We define $$\prob{\adv(\secparam)} := \prob{\adv(\secparam) = 1}$$ as a
% shorthand for $\adv$'s success probability.
% Security property
Security holds if no adversary has non-negligible advantage.
% Security reduction
The proof is achieved by reducing the winning of the game to some problem that is assumed to be hard.

% Sequence of games
We perform security reductions using a \emph{sequence of games}.
The first game is the original security experiment provided by the security definition.
Each subsequent game is equal to the previous game except for some small well-defined change for which we argue that it does only negligibly influence adversarial advantage.
The last game then has a special and easy to verify property, \eg the adversary has no advantage over a blind guess.
Only negligible change in advantage between subsequent games implies only negligible change in advantage between the first and the last game, which concludes the reduction.

\subsection{Homomorphic Authenticated Symmetric Encryption (HASE)}

\begin{figure}[bt]
\centering
\includegraphics[width=\columnwidth]{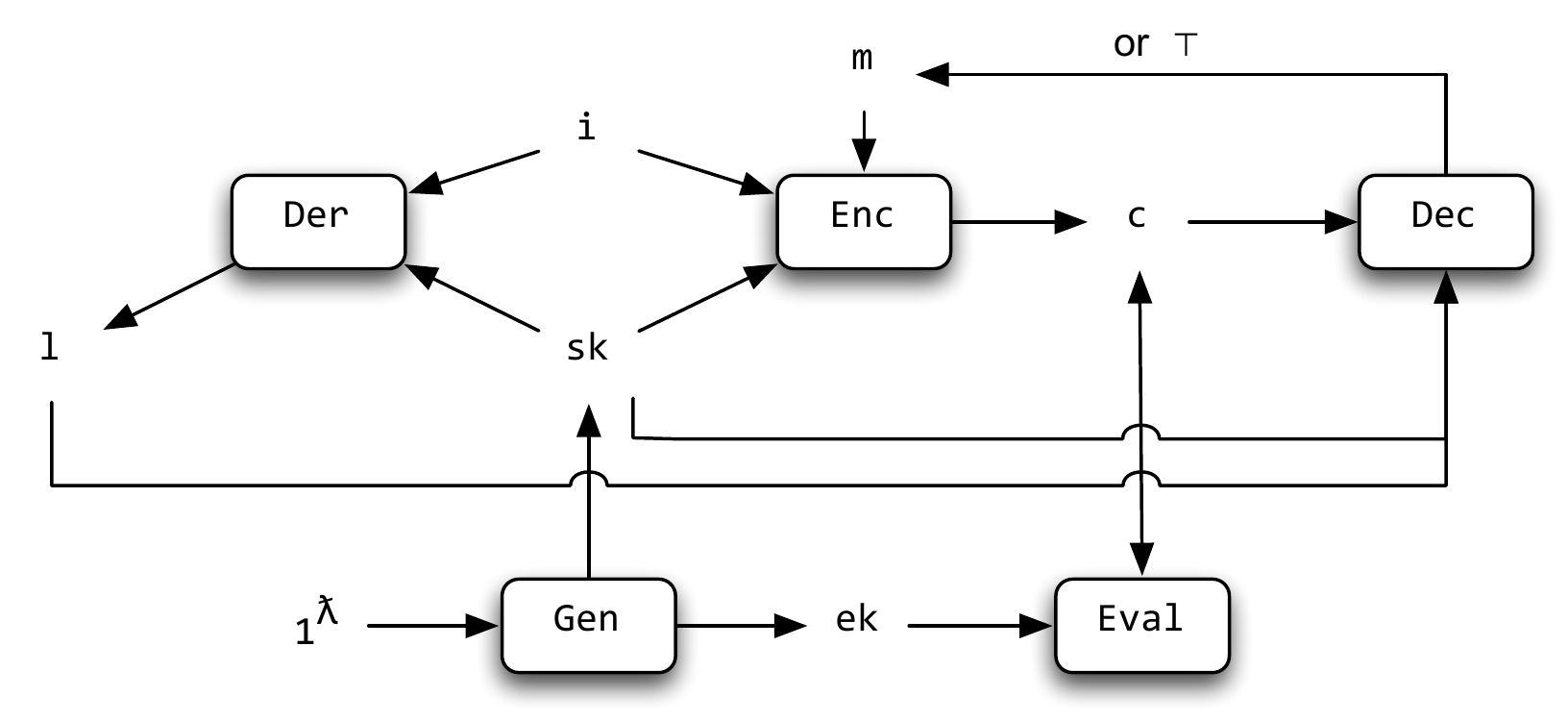}
\caption{HASE Overview}
\label{fig:scheme}
\end{figure}

In this section, we discuss the syntax, correctness and security of a HASE scheme.
For security we define confidentiality in terms of indistinguishability and authenticity in terms of unforgeability.
Indistinguishability of HASE schemes (HASE-IND-CPA) is defined as an adaptation of the commonly used IND-CPA security definition for symmetric encryption schemes \cite{Katz:2014:IMC:2700550}.
Unforgeability of HASE schemes (HASE-UF-CPA) is based on the common \emph{unforgeable encryption} definition \cite{Katz:2014:IMC:2700550}.

\label{sec:syntax}
\begin{defn}[HASE Syntax]
A HASE scheme is a tuple of PPT algorithms $\Angle{\kgen, \enc, \eval, \der, \dec}$ such that:
\begin{itemize}
 \item
  The \emph{key-generation algorithm \kgen} takes the security parameter $\secparam$ as input and outputs a key pair $\langle ek, sk \rangle$ consisting of a \emph{public evaluation key} $ek$ and a \emph{secret key} $sk$.
  The evaluation key implicitly defines a commutative plaintext group $(\Msp, \oplus)$, a commutative ciphertext group $(\Csp, \otimes)$ and a commutative label group $(\Lsp, \diamond)$.

 \item The \emph{encryption algorithm \enc} takes a secret key $sk$, a plaintext message $m \in \Msp$ and an identifier $i \in \Isp$ as input and outputs a ciphertext $c \in \Csp$.

 \item The \emph{evaluation algorithm \eval} takes an evaluation key $ek$ and a set of ciphertexts $C \subseteq \Csp$ as input and outputs a ciphertext $\hat{c} \in \Csp$.

 \item The deterministic \emph{label derivation algorithm \der} takes a secret key $sk$ and a set of identifiers $I \subseteq \Isp$ as input and outputs a secret label $l \in \Lsp$.

 \item The deterministic \emph{decryption algorithm \dec} takes a secret key $sk$, a ciphertext $c \in \Csp$ and a secret label $l \in \Lsp$ as input and outputs a plaintext message $m \in \Msp$ or $\bot$ on decryption error.
\end{itemize}
\end{defn}

An overview of all operations involved in our HASE encryption scheme is provided in Figure~\ref{fig:scheme}.

\begin{defn}[HASE Correctness]
Let $\Pi$ be a HASE scheme consisting of five algorithms as described above.
Furthermore, let $$S := \{ (m, i) : m \in \Msp, i \in \Isp \}$$ be a set of plaintexts with unique identifiers and let $I := \pi_2(S)$ be the set of identifiers in $S$.
We say that $\Pi$ is \emph{correct} if for any honestly generated key pair $\Angle{ek, sk} \gets \Pi.\kgen(\secparam)$ and any set of ciphertexts $$C := \{ c : c \gets \Pi.\enc(sk, m, i) : (m, i) \in S \}$$ it holds that
 $$
 \Pi.\dec(sk, \Pi.\eval(ek, C), \Pi.\der(sk, I)) =  \bigoplus_{(m, i) \in S} m
 $$
 except with negligible probability over $\Angle{ek, sk}$ output by $\allowbreak \Pi.\kgen(\secparam)$.
\end{defn}

% TODO: We might actually want to remove this altogether.
Note that when $S$ is a singleton set, \ie we have
 $S := \{ (m, i) \}$,
 $I := \{ i \}$,
 $l := \Pi.\der(sk, I)$ and
 $C := \{ c \}$ for some $c \gets \Pi.\enc(sk, m, i)$,
 we expect
 $\Pi.\eval(C)$ to return $c$ and
 the last equation in the above definition boils down to
 $\Pi.\dec(sk, c, l) = m$.

 % [p.~74]
%The experiment proceeds as follows:
%
%The challenger invokes $\Pi.\kgen$ to generate a key pair and provides the
% evaluation key to the adversary.
%The adversary is given access to an encryption oracle which on input a message
% and an identifier outputs a corresponding ciphertext produced using $\Pi.\enc$
% and the secret key.
%The oracle enforces uniqueness of the identifier by keeping track of all
% queried plaintext-identifier pairs.
%It rejects any encryption query involving a previously used identifier.
%
%At some point the adversary outputs two plaintexts, a \emph{challenge
% identifier} and state $\state$.
%The adversary loses the game if the challenge identifier was used in a previous
% oracle query.
%The challenger flips a fair coin and depending on the output of the coin flip
% either selects the first or the second plaintext as the
% \emph{challenge plaintext}.
%By invoking $\Pi.\enc$ on the secret key, the challenge plaintext and the
% challenge identifier, the challenger produces the \emph{challenge ciphertext}.
%The challenge ciphertext, state $\state$ and access to the oracle described
% above is given to the adversary.
%At some point the adversary outputs a bit indicating which of the two chosen
% plaintexts was encrypted by the challenger.
%The adversary wins if the returned bit indicates the correct plaintext.
%
%The formal definition is as follows.

\begin{defn}[HASE-IND-CPA]
\label{def:hase-ind-cpa}
A HASE scheme $\Pi$ has \emph{indistinguishable encryptions under a
 chosen-plaintext attack}, or is \emph{CPA-secure}, if for all PPT adversaries
 $\adv$ there is a negligible function $\negl$ such that
\begin{align*}
 \advantage{\indcpa}{\adv,\Pi} &:=
 \left|
  \prob{\mathrm{ExpHASE}_{\adv,\Pi}^{\indcpa}(\secpar) = 1} - \dfrac{1}{2}
 \right| \\
 & \leq \negl
\end{align*}
The experiment is defined as follows:
\begin{pcvstack}[center]
\procedure[space=auto]
 {$\mathrm{ExpHASE}_{\adv,\Pi}^{\indcpa}(\secpar)$}{
 S := \{ \} \\
 \Angle{ek, sk} \gets \Pi.\kgen (\secparam) \\
 \Angle{m_0, m_1, i, \state} \gets \adv^{E_{sk}}(\secparam, ek) \\
 \pcif i \in \pi_2(S) \pcthen \\
  \pcreturn 0 \\
 \pcelse \\
  b \sample \bin \\
  S := S \cup \{ (m_b, i) \} \\
  c \gets \Pi.\enc(sk, m_b, i) \\
  b' \gets \adv^{E_{sk}}(\secparam, c, \state) \\
  \pcreturn b = b'
}
\pchspace
\procedure[space=auto]
 {$E_{sk}(m, i)$}{
 \pcif i \in \pi_2(S) \pcthen \\
  \pcreturn \bot \\
 \pcelse \\
  S := S \cup \{ (m, i) \} \\
  c \gets \Pi.\enc(sk, m, i) \\
  \pcreturn c
}
\end{pcvstack}
\end{defn}

The differences to symmetric encryption are as follows:
We removed the explicit requirement that $m_0$ and $m_1$ need to be of the same bit length because we consider plaintexts to be elements of some group $(\Msp, \oplus)$ rather than bit strings.
The extra identifier parameter to the encryption algorithm is incorporated by allowing the adversary to submit an additional identifier argument to encryption oracle queries.
Additionally, the adversary is allowed to pick the identifier used for the encryption of the challenge plaintext $m_b$.

\begin{defn}[HASE-UF-CPA]
\label{def:hase-uf-cpa}
A HASE scheme $\Pi$ is \emph{unforgeable under a chosen-plaintext attack}, or
 just \emph{unforgeable}, if for all PPT adversaries $\adv$ there is a
 negligible function $\negl$ such that
\begin{align*}
\advantage{\ufcpa}{\adv,\Pi} &:=
 \prob{\mathrm{ExpHASE}_{\adv,\Pi}^{\ufcpa}(\secpar) = 1} \\
 & \leq \negl
\end{align*}
 with the experiment defined as follows:
\begin{pcvstack}[center]
\procedure[space=auto]
 {$\mathrm{ExpHASE}_{\adv,\Pi}^{\ufcpa}(\secpar)$}{
 S := \{ \} \\
 \Angle{ek, sk} \gets \Pi.\kgen (\secparam) \\
 (c, I) \gets \adv^{E_{sk}}(\secparam, ek) \\
 l := \Pi.\der(sk, I) \\
 m := \Pi.\dec(sk, c, l) \\
 \tilde{m} := \bigoplus_{(m', i) \in S, i \in I} m' \\
 \pcreturn m \neq \bot \land m \neq \tilde{m}
}
\pchspace
\procedure[space=auto]
 {$E_{sk}(m, i)$}{
 \pcif i \in \pi_2(S) \pcthen \\
  \pcreturn \bot \\
 \pcelse \\
  S := S \cup \{ (m, i) \} \\
  c \gets \Pi.\enc(sk, m, i) \\
  \pcreturn c
}
\end{pcvstack}
\end{defn}

The challenger invokes $\Pi.\kgen$ to generate a key pair and provides the
 evaluation key to the adversary.
The adversary is given access to an encryption oracle which on input a message
 and an identifier outputs a corresponding ciphertext produced using $\Pi.\enc$
 and the secret key.
The oracle enforces uniqueness of the identifier by keeping track of all
 queried plaintext-identifier pairs.
It rejects any encryption query involving a previously used identifier.

At some point the adversary outputs a ciphertext and a set of
 identifiers.
The adversary wins if the returned ciphertext is a valid forgery with respect
 to the set of identifiers.
This is the case if and only if two conditions are met.
First, the ciphertext has to successfully decrypt under the label derived from
 the set of identifiers returned by the adversary.
Second, the resulting plaintext must be different from the plaintext resulting
 from the application of the plaintext operation to the set of plaintexts
 corresponding to the set of identifiers returned by the adversary.

% TODO: Add comparison to referenced security definition here?

\section{Data Flow Authentication (\dfauth)}
\label{sec:coed}
We introduce \dfauth using an example.
Consider the following excerpt from a Java program:

\begin{lstlisting}
a = b + c;
d = a * e;
if (d > 42)
    f = 1;
else
    f = 0;
\end{lstlisting}

First, \dfauth performs a conversion to single static assignment (SSA) form~\cite{alpern1988detecting}: assign each variable at exactly one code location; create atomic expressions; introduce fresh variables if required.
In the example, \dfauth changes the program to the following:

\begin{lstlisting}
a = b + c;
d = a * e;
d1 = d > 42;
if (d1)
    f1 = 1;
else
    f2 = 0;
f = phi(f1,f2);
\end{lstlisting}

As usual in SSA, \verb!phi! is a specially interpreted merge function that combines the values of both assignments to \verb!f!, here denoted by \verb!f1! and \verb!f2!.

\dfauth then performs a type inference similar to JCrypt \cite{DonMil16} and AutoCrypt \cite{ACCCS13}.
As a result of this inference, each variable and constant is assigned an encryption type of $\{ add, \allowbreak mul, \allowbreak cmp \}$.
At runtime, each constant and variable value will be encrypted according to the appropriate type.
Our HASE encryption implements multiplicative homomorphic encryption $mul$ and its operations directly, while it implements additive homomorphic encryption $add$ using exponentiation.
Comparisons $cmp$ are implemented in the Trusted Module.
Our experiments show that this is more efficient than performing the comparison in the program space using conversion to searchable or functional encryption.
An attacker observing user space will hence only see encrypted variables and constants, but can observe the control flow.
Actual data values are hidden from the attacker.
%TODO Do we show this, did other show this? (then cite)
%The leakage, however, \ie the result of the comparison, is the same.

Combinations of multiple operations, however, require additional work.
Every time a variable is encrypted in one encryption type (\eg additive), but is later used in a different one (\eg multiplicative), \dfauth must insert a conversion.
The resulting program in our running example looks as follows:

\begin{lstlisting}[caption=Example as executed on the server, label=lst:example]
a = b + c;
a1 = convertToMul(a, "a1");
d = a1 * e;
d1 = convertToCmpGT42(d, "d1");
if (d1)
    f1 = 1;
else
    f2 = 0;
f = phi(f1,f2);
\end{lstlisting}

The first conversion is necessary because the variable \verb!a! must be converted from additive to multiplicative homomorphic encryption.
The resulting re-encrypted value is stored in \verb!a1!.
For security reasons, the decryption performed by the conversion routine must be sensitive to the variable identifier it is assigned to.
A unique label must be introduced to make the decryption routine aware of the code location.
\dfauth can use the left-hand-side variable's identifier (\verb!"a1"! in this example), because it introduced unique names during SSA conversion.
Using this variable identifier, the conversion routine can retrieve the corresponding label of the HASE encryption stored in the memory protected by the Trusted Module.

Any branch condition is also treated as a conversion that leaks the result of the condition check.
In the example, \dfauth introduces the variable \verb!d1! to reflect this result:
\begin{lstlisting}[firstnumber=4]
d1 = convertToCmpGT42(d, "d1");
\end{lstlisting}
% In this special case we compare \verb!d! to the constant $42$.
To simplify the exposition, we assume that our compiler inlines this comparison into a special routine \verb!convertToCmpGT42!.
In the general case, a binary comparison on two variables \verb!x! and \verb!y! would result in a call to a routine \verb!convertToCmp(x,y,"z")!.
We show the full algorithm in Listing~\ref{lst:comp} in Section~\ref{sec:impl} which is generic for all comparisons and in case of comparison to a constant looks this constant up in an internal table protected by the Trusted Module. %\todo{EB: Should we show Listing~\ref{lst:comp} here?}
We need to protect constants in comparisons, since if they were part of the program text, they could be modified by the adversary.
%\todo{EB: Wir sollten beschreiben warum wir einen paramLookup machen.}

As mentioned before, the security challenge of such conversions to $cmp$ is that they leak information about the encrypted variables, and particularly that active adversaries that can modify the control and data flow can exploit those leaks to restore the variables' plaintext.
%This is why previous approaches such as AutoCrypt disallow conversions from homomorphic encryption to comparisons entirely.
%Of course, this very much restricts the programs one can run with encryption.
%
In this paper, we thus propose to restrict the data flow using \dfauth.
It allows such conversions in a secure way by enforcing that encrypted variables can be decrypted only along the program's original data flow.
%TODO EB: Should we not call this data-flow authentication?
The approach comprises two steps.
First, happening at compile time, for each conversion \dfauth precomputes the $\der$ function on the operations in the code.
In the conversion \verb!convertToMul(a, "a1")! (at line~2 in our example), \dfauth computes the label
$$l2 = \der(sk, \{\verb!"b"!, \verb!"c"!\})$$
and in the conversion at line~4
$$l4 = \der(sk, \{\verb!"a1"!, \verb!"e"!\})$$
Here the second argument to $\der$ is the multi-set of variable identifiers involved in the unique computation preceding the conversion.
(We use a multi-set and not a vector, because all our encrypted operations are commutative.)
The compiler computes labels for all variables and constants in the program.

At runtime the computed labels as well as the secret key $sk$ are both kept secret from the attacker, which is why both are securely transferred  to, and stored in, the Trusted Module during the setup phase.
The Trusted Module registers the secret labels under the respective identifier, for example, associating the label $l4$ with the identifier \verb!"d1"!.

All conversion routines run within the Trusted Module.
They retrieve a secret label for an identifier with the help of a \verb!labelLookup(id)! function.
In particular, when the program runs and a conversion routine is invoked, the Trusted Module looks up and uses the required labels for decryption.
In the example at line~4, the call to \verb!convertToCmpGT42! internally invokes the decryption operation \verb!Dec(sk, d, l4)! using secret label \verb!l4! retrieved for variable identifier \verb!"d1"!:
% According to the security definition of our modified, authenticated Elgamal style encryption scheme, this decryption only succeeds if the program follows the control and data flow.
% Deviations result in incorrect labels and are detected.
% This way one can prevent the adversary from deviating from the program.
% Note that the same security holds for the conversion to comparison including leaking the correct result.

\begin{lstlisting}
convertToCmpGT42(d, "d1") {
  l4 = labelLookup("d1");
  x = Dec(sk, d, l4);
  if (x == fail)
    stop;
  return (x > 42);
}
\end{lstlisting}
%TODO EB: can we write this that way? Is the unencrypted result returned?
% How is "a1" part of the Dec-Operation?

Note that in this scheme, the Trusted Module returns the result of the comparison in the clear.
In this case, however, leaking the branch decision is \emph{secure}, as the HASE encryption scheme guarantees that any active attack that would yield the adversary a significant advantage will be reliably detected.

Let us assume an attacker that attempts to modify the program's data or control flow to leak information about the encrypted plaintexts, for instance, using a binary search like we described in the introduction.
The attacker is not restricted to the compiled instructions in the program, and can also try to ``guess'' the result of cryptographic operations as the adversary in experiment $\mathrm{ExpHASE}_{\adv,\Pi}^{\indcpa}$.
This modification to binary search can only succeed if the decryption operations $\dec$ in \verb!convertToCmpGT42! (or other conversion routines) succeed.
% %TODO EB: Can one find out something baout the plaintext through a failing decryption? (side channel)
The adversary can minimize the $\dec$ operations, \eg by not introducing new calls to conversion routines, but given the scheme defined above, any attempt to alter the data flow on encrypted variables will cause $\dec$ to fail:
Assume that an attacker inserts code in Listing~\ref{lst:example} to search for the secret value \verb!d! (resulting code shown in Listing~\ref{lst:examplemod}).
We only use this code to illustrate potential attacks and ignore the fact that the attacker would need access to the encrypted constants (\verb!2^i!) and needs to guess the result of the homomorphic addition operation on the ciphertexts.
However, given these capabilities, the attacker could try to observe the control flow -- simulated by our statement \verb!leak(f)! -- which then would in turn leak the value of \verb!d!.
\begin{lstlisting}[caption=Example modified by the attacker, label=lst:examplemod]
a = b + c;
a1 = convertToMul(a, "a1");
g1 = a1 * e;                     //changed
for(i = n..1) {                  //inserted
  g = phi(g1, g3);               //inserted
  d = g + 2^i;                   //inserted
  d1 = convertToCmpGT42(d, "d1");
  if (d1) {
      f1 = 1;
      g2 = g3 - 2^i;             //inserted
  } else
      f2 = 0;
  g3 = phi(g, g2);               //inserted
  f = phi(f1,f2);
  leak(f);                       //inserted
}
\end{lstlisting}
%\todo{EB: Fehlt nicht fÃŒr die Berechnung von $2^i$ auch noch eine Konvertierung?}
This code will only execute if each variable decryption succeeds, but decryption for instance of \verb!d1! will succeed only if it was encrypted with the same label \verb!l4! that was associated with \verb!d1! at load time.
Since the Trusted Module keeps secret the labels themselves, and also the secret key $sk$ to compute labels, the attacker cannot possibly forge the required label at runtime.
Moreover, in the attacker-modified program, the encryption must fail due to the altered data dependencies: in the example, the input \verb!d! to \verb!convertToCmpGT42! has now been derived from \verb!g3! and \verb!i! instead of \verb!a1! and \verb!e!, which leads to a non-matching label for \verb!d!.
In result, the decryption in the conversion routine \verb!convertToCmpGt42! will fail (in this specific attack even at the first comparison) and stop program execution before any unintended leakage can occur.
%\todo{EB: won't we fail here already at convertToPlus?}

\label{sec:ie}
More generally, the way in which we derive labels from data-flow relationships enforces a notion of \emph{interference equivalence}.

\begin{defn}[Non-interference]
Generally, a program $P$ is said to be \emph{non-interferent}, if applied to two different memory configurations $M_1, M_2$ that are equal w.r.t.\ their low, \ie unclassified memory locations, $M_1, =_L M_2$ for short, then also the resulting memory locations must show such low-equivalency: $P(M_1), =_L P(M_2)$. Non-interference holds if and only if there is no information flow from high, \ie classified values to low memory locations. While this is a semantic property, previous research has shown that one can decide non-interference also through a structural analysis of programs, through so-called Program Dependency Graphs (PDGs) that capture the program's control and data flow~\cite{WasLohSnelting}. In this view, a program is \emph{non-interferent} if the PDG is free of paths from high to low memory locations.
\end{defn}

\begin{defn}[Non-interference under declassification]
In the setting considered in this paper one must assume that the executed program \emph{before encryption} already shows interference for some memory locations, \eg because the program is, in fact, intended to declassify some limited information. Let us denote by $M\downarrow C$ a projection of memory configuration $M$ onto all (classified) memory locations $C$ that are not declassified that way. Then even in the setting here it holds for any program $P$ and any memory configurations $M_1, M_2$ that $P(M_1\downarrow C) =_L P(M_2\downarrow C)$.
\end{defn}

We next state and prove the main theorem of our construction: that any program that an attacker can produce, and that would lead to the same computation of labels (and hence decryptable data) as the original program, cannot produce any more information flows than the original program.

\begin{thm}[Preservation of non-interference]
\label{thm:noninterf}
Let us denote by $tr$ a program transformation conducted by the attacker, \eg the transformation explained above, which inserted a binary search. Then, by construction, it holds that:
\begin{eqnarray*}
 &&\forall M_1, M_2, tr: P(M_1\downarrow C) =_L P(M_2\downarrow C)\\
 &&\rightarrow (tr(P))(M_1\downarrow C) =_L (tr(P))(M_2\downarrow C)
\end{eqnarray*}
\end{thm}
In other words: disregarding the explicitly declassified information within $C$, also the transformed program does not leak any additional information.

\begin{proof}[Proof by Contradiction]
Assume that Theorem~\ref{thm:noninterf} did not hold. Then there would exist a transformation $tr$ that would cause the transformed program $tr(P)$ to compute values in at least one low memory location despite low-equivalent inputs. But this is impossible, as any such transformation would necessarily have to insert additional PDG-edges, destroying the label computations, and hence invalidating the decryptions in our HASE encryption scheme.
%TODO we seem that have termination-insensitive non-interference only. is that an issue?
\end{proof}

\paragraph*{Result Verification}

Note that the client can verify the result of the computation using a simple check on the variable's label -- just as the conversion routine does.
The result is just another variable, which albeit not being converted, can be checked for correct data flow computation.
That way, a client can ensure that it receives a valid output of the program.

\section{HASE Constructions}
\label{sec:hase}
In this section, we provide two constructions of HASE schemes: one homomorphic with respect to multiplication (Multiplicative HASE) and another with respect to addition (Additive HASE), on integers.

\subsection{Constructions}
Our first construction is based on the renowned public-key encryption scheme due to Elgamal \cite{Elgamal}.
We do not make use of the public-key property of the scheme, but extend ciphertexts with a third group element working as a homomorphic authenticator.

\begin{cnst}[Multiplicative HASE]
\label{cnst:elghase}
Let $\mathcal{G}$ be a group generation algorithm and let the DDH problem
 be hard relative to $\mathcal{G}$ (cf.~Definition \ref{def:ddh}).
Define a HASE scheme using the following PPT algorithms:
 \begin{itemize}
  \item \kgen:
   on input $\secparam$ obtain
    $\Angle{\mathbb{G}, q, g} \gets \mathcal{G}(\secparam)$.
   For a pseudorandom function family
    $$H : \mathcal{K} \times \Isp \rightarrow \mathbb{G}$$
   choose $k \sample \mathcal{K}$.
   Choose $a, x, y \sample \mathbb{Z}_q$ and compute $h := g^x$, $j := g^y$.
   The evaluation key is $\elghaseek$, the secret key is $\elghasesk$.
   The plaintext group is $(\Msp, \oplus) := (\mathbb{G}, \cdot)$ where
    $\cdot$ is the group operation in $\mathbb{G}$.
   The ciphertext group is $(\mathbb{G}^3, \otimes)$ where we define $\otimes$
    to denote the component-wise application of $\cdot$ in $\mathbb{G}$.
   The label space is $(\mathbb{G}, \cdot)$.

  \item \enc:
   on input a secret key $sk = \elghasesk$, a message $m \in \mathbb{G}$ and an
    identifier $i \in \Isp$.
   Choose $r \sample \mathbb{Z}_q$ and obtain the label $l = H(k, i)$.
   Compute
    $u := g^r$,
    $v := h^r \cdot m$ and
    $w := j^r \cdot m^a \cdot l$.
   Output the ciphertext $\Angle{u, v, w}$.

  \item \eval:
   on input an evaluation key $\elghaseek$ and a set of ciphertexts
    $C \subseteq \Csp$ compute the ciphertext
    $$
     c := \bigotimes_{c' \in C} c'
    $$
   and output $c$.

  \item \der:
   on input a secret key $\elghasesk$ and a set of
    identifiers $I \subseteq \Isp$ compute the label
    $$
     l := \prod_{i \in I} H(k, i)
    $$
   and output $l$.
   Note that here $\Pi$ denotes the repeated application of the group operation
    $\cdot$ in $\mathbb{G}$.

  \item \dec:
   on input a secret key $\elghasesk$, a ciphertext $c = \Angle{u, v, w}$
    and a secret label $l \in \mathbb{G}$.
   First compute
    $m := {u}^{-x} \cdot v$, then
    $t := {u}^y \cdot m^a \cdot l$.
   If $t$ equals $w$ output $m$, otherwise output $\bot$.
 \end{itemize}
\end{cnst}

It is well known that the Elgamal encryption scheme is homomorphic with regard
 to the group operation in $\mathbb{G}$.
As can be easily seen, this property is inherited by our construction.
For the original Elgamal scheme, $\mathbb{G}$ is most commonly instantiated
 either as $\mathbb{G}_q$, the $q$-order subgroup of quadratic residues of
 $\mathbb{Z}_{p}^{*}$ for some prime $p = 2q + 1$ (with $q$ also prime), or
 as an elliptic curve over some $q$-order finite field.
In the latter case, the group operation is elliptic curve point addition and
 the ability to perform it in a homomorphism serves no useful purpose in our
 context.
Instantiating $\mathbb{G}$ as $\mathbb{G}_q$ on the other hand enables
 homomorphic multiplication on the integers.
%TODO: Explain that plaintexts need to be encoded as group elements.
% Maybe mention this in the implementation.

Our second construction supports homomorphic integer addition and is obtained
 by applying a technique proposed by Hu et al.~\cite{CRT-ELG} to the construction
 presented above.
The basic idea is to consider plaintexts to be element of $\mathbb{Z}_q$
 instead of $\mathbb{G}$ and to encrypt a given plaintext $m$ by first raising
 the generator $g$ to the power of $m$ and then encrypting the resulting group
 element in the usual way.
In detail, this means computing ciphertexts of the form $\Angle{g^r, h^r g^m}$
 rather than $\Angle{g^r, h^r m}$.
To see that the resulting scheme is homomorphic with regard to addition on
 $\mathbb{Z}_q$, consider what happens when the group operation is applied
 component-wise to two ciphertexts:
$$
\Angle{g^{r_1} \cdot g^{r_2}, h^{r_1} g^{m_1} \cdot h^{r_2} g^{m_2}} =
\Angle{g^{r_1+r_2}, h^{r_1+r_2} \cdot g^{m_1+m_2}}
$$
Unfortunately, decryption now involves computing discrete logarithms with
 respect to base $g$, which must be difficult for sufficiently large exponents
 in order for the DDH problem to be hard relative to $\mathcal{G}$.
Hu et al.~keep exponents small enough for discrete logarithm algorithms to
 terminate within reasonable time despite their exponential asymptotic runtime.
They do so by unambiguously decomposing plaintexts $m$ into $t$ smaller
 plaintexts $m_e$ ($e \in [1, t]$) via means of the Chinese remainder theorem
 (CRT) and then encrypting each $m_e$ separately.
Although doing so increases the ciphertext size roughly by a factor of $t$ in
 comparison to Construction~\ref{cnst:elghase}, this drawback can be
 compensated by instantiating $\mathbb{G}$ as an elliptic curve group
 since the homomorphic operation is on $\mathbb{Z}_q$ rather than $\mathbb{G}$.
At a comparable security level, group elements of elliptic curves can be
 represented using a fraction of bits \cite{KeyLength}.

We provide the full details of our \emph{Additive HASE} construction in the
 following.
Note how the authenticator only requires constant (\ie independent of $t$)
 ciphertext space and can be verified without discrete logarithm computation.
Although we consider instantiating $\mathbb{G}$ as an elliptic curve group,
 we keep writing the group operation multiplicatively
 (cf.~Section~\ref{sec:prel:not}).

\begin{cnst}[Additive HASE]
\label{cnst:ceghase}
Let $\mathcal{G}$ be a group generation algorithm as before.
Define a HASE scheme using the following PPT algorithms and the $\eval$
 algorithm from Construction~\ref{cnst:elghase}:
 \begin{itemize}
  \item \kgen:
   on input $\secparam$ obtain
    $\Angle{\mathbb{G}, q, g} \gets \mathcal{G}(\secparam)$.
   For a pseudorandom function family
    $$H : \mathcal{K} \times \Isp \rightarrow \mathbb{Z}_q$$
   choose $k \sample \mathcal{K}$.
   Choose $\{d_1, \ldots, d_t\} \subset \mathbb{Z}^{+}$ such that
    $d := \prod_{e=1}^t d_e < q$ and $\forall e \neq j : \gcd(d_e, d_j) = 1$.
   Define $D := \ceghaseski$.
   Choose $a, x, y \sample \mathbb{Z}_q$ and compute $h := g^x$, $j := g^y$.
   The evaluation key is $\ceghaseek$, the secret key is $\ceghasesk$.
   The plaintext group is $(\Msp, \oplus) := (\mathbb{Z}_d, +)$.
   The ciphertext group is $(\mathbb{G}^{2(t+1)}, \otimes)$ where $\otimes$
    denotes the component-wise application of $\cdot$ in $\mathbb{G}$.
   The label space is $(\mathbb{G}, \cdot)$.

  \item \enc:
   on input a secret key $sk = \ceghasesk$, a message $m \in \mathbb{Z}_d$ and
    an identifier $i \in \Isp$.
   Obtain the label $l := H(k, i)$.

   For $e := 1, \ldots, t$:
   \begin{itemize}
    \item Compute $m_e := m \bmod d_e$.
    \item Choose $r_e \sample \mathbb{Z}_q$
    \item Compute $u_e := g^{r_e}$
    \item Compute $v_e := h^{r_e} \cdot g^{m_e}$
   \end{itemize}
   Choose $r \sample \mathbb{Z}_q$.
   Compute $s := g^r$ and $w := j^r \cdot g^{m^a} \cdot l$.
   Output the ciphertext $\ceghasect$.

  \item \der:
   on input a secret key $\ceghasesk$ and a set of
    identifiers $I \subseteq \Isp$ compute the label
    $$
     l := \prod_{i \in I} g^{H(k, i)}
    $$
   and output $l$.

  \item \dec:
   % Elgamal Decrypt, dlog, CRT Recompose
   on input a secret key $\ceghasesk$, a ciphertext $\ceghasect$ and a secret
    label $l \in \mathbb{G}$.
   Parse $D = \ceghaseski$.
   First compute $m_e := \log_g(v_e {u_e}^{-x})$ for $e = 1, \ldots, t$,
    then recover
   $$
   m := \sum_{e=1}^t m_e \frac{d}{d_e}
       \left( \frac{d}{d_e}^{-1} \bmod d_e \right)
       \bmod d
       \text{.}
   $$
   If $s^y \cdot g^{m^a} \cdot l = w$ then output $m$, else output $\bot$.
   Note that $\log_g$ denotes the discrete logarithm with respect to base $g$.
 \end{itemize}
\end{cnst}

\subsection{Security Reductions}
\label{sec:secred}
We define the assumptions and security the HASE schemes provide under these assumptions.
However, in order to maintain the page limit proofs are deferred to Appendix~\ref{app:secred}.

\begin{defn}[Pseudorandom Function]
\label{def:prf}
Let $X$ and $Y$ be two finite sets and denote the set of all functions from $X$
 to $Y$ as $\mathcal{F}$.
We say that an efficiently computable keyed function $F : \mathcal{K} \times X
 \rightarrow Y$ with keyspace $\mathcal{K}$ is a pseudorandom function (PRF),
 if for all PPT algorithms $\adv$ there is a negligible function $\negl$ such
 that:
\begin{align*}
\left|
 \prob{\adv^{F(k, \cdot)}(\secparam) = 1} -
 \prob{\adv^{f(\cdot)}(\secparam) = 1}
\right| \leq \negl
\end{align*}
where the first probability is taken over $k \sample \mathcal{K}$ and the
 second probability is taken over $f \sample \mathcal{F}$.
\end{defn}

% \cite{Katz:2014:IMC:2700550}[p.~321]
\begin{defn}[DDH Problem \cite{Katz:2014:IMC:2700550}]
\label{def:ddh}
Let $\mathcal{G}$ be a PPT algorithm taking $\secparam$ as input and outputting
 $\Angle{\mathbb{G}, q, g}$ where $\mathbb{G}$ is a description of a cyclic
 group, $q$ is the order of $\mathbb{G}$ and $g$ is a generator of $\mathbb{G}$.
We say that \emph{the Decisional Diffie-Hellman (DDH) problem is hard relative
 to $\mathcal{G}$} if for all PPT algorithms $\adv$ there is a negligible
 function $\negl$ such that:
\begin{align*}
\Big|
 &\prob{\adv(\mathbb{G}, q, g, g^\alpha, g^\beta, g^\gamma) = 1} - \\
 &\prob{\adv(\mathbb{G}, q, g, g^\alpha, g^\beta, g^{\alpha \beta}) = 1}
\Big| \leq \negl
\end{align*}
where in each case the probabilities are taken over the experiment in which
 $\mathcal{G}(\secparam)$ outputs $\Angle{\mathbb{G}, q, g}$, and then
 $\alpha, \beta, \gamma \sample \mathbb{Z}_q$.
\end{defn}

\begin{thm}[Multiplicative HASE-IND-CPA]\label{thm:elg-hase-ind-cpa}
Let $\Pi$ be Construction~\ref{cnst:elghase}.
If the DDH problem is hard relative to $\mathcal{G}$ and $H$ is a PRF as
 described in $\Pi.\kgen$, then $\Pi$ is CPA-secure.
\end{thm}

\begin{thm}[Multiplicative HASE-UF-CPA]\label{thm:elg-hase-uf-cpa}
Let $\Pi$ be Construction~\ref{cnst:elghase}.
If $H$ is a PRF as described in $\Pi.\kgen$, then $\Pi$ is unforgeable.
\end{thm}

% Note that \Pi is different and as such \Pi.\kgen and as such H.
\begin{lem}[Additive HASE-IND-CPA]
\label{lem:ceg-hase-ind-cpa}
Let $\Pi$ be Construction~\ref{cnst:ceghase}.
If the DDH problem is hard relative to $\mathcal{G}$ and $H$ is a PRF as
 described in $\Pi.\kgen$, then $\Pi$ is CPA-secure.
\end{lem}

\begin{lem}[Additive HASE-UF-CPA]
\label{lem:ceg-hase-uf-cpa}
Let $\Pi$ be Construction~\ref{cnst:ceghase}.
If $H$ is a PRF as described in $\Pi.\kgen$, then $\Pi$ is unforgeable.
\end{lem}

\begin{figure*}[t]
\centering
\includegraphics[width=\textwidth]{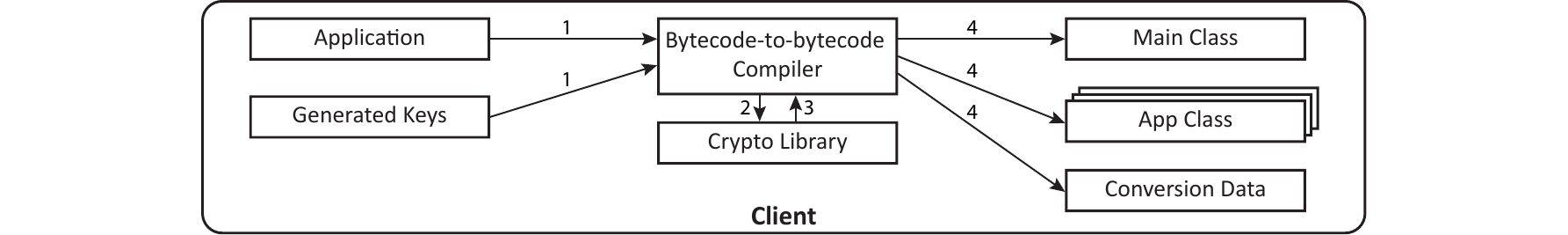}
\caption{
 Application Transformation during setup phase. See Section~\ref{sec:impl-setup} for a full description.
}
\label{fig:impoverview-setup}
\end{figure*}

\section{Implementation}
\label{sec:impl}
In this section, we present implementation details %of a prototype that is 
used in the subsequent evaluation section.
Recall from Section~\ref{sec:adversary} that we consider a scenario between a trusted \emph{client} and an untrusted \emph{cloud server} (which has a Trusted Module) and we distinguish two phases of the outsourced computation: \emph{setup} and \emph{runtime}.

\subsection{Setup Phase}
\label{sec:impl-setup}

In Figure~\ref{fig:impoverview-setup}, we present the setup phase in more detail.% than in the introduction section.

\paragraph*{Compilation} 
First, the client translates any Java bytecode program to a bytecode program running on encrypted data.
To start, the client generates a set of cryptographic keys for our HASE encryption.
It then uses our \emph{Bytecode-to-Bytecode Compiler} to transform an application (in the form of Java bytecode) using the \emph{Generated Keys} (1).
Our compiler is based on \emph{Soot}, ``a framework for analyzing and transforming Java and Android Applications'' \cite{SOOT}.
%More specifically, we implemented a \texttt{SceneTransformer} that uses the facilities provided by Soot to analyse and manipulate aspects of Java programs.

Our implementation uses a \emph{Crypto Library} to encrypt program constants and choose variable labels (2-3).
The crypto library contains implementations of all cryptographic algorithms, including our own cryptographic algorithms from Section~\ref{sec:hase}.
It implements the PRF used for the authentication labels as HMAC-SHA256 \cite{RFC4634}.
For the group operations in Multiplicative HASE we use the MPIR (Multiple Precision Integers and Rationals) library \cite{MPIR} for large integer arithmetic.
Additive HASE operates on the elliptic curve group provided by \verb+libsodium+ \cite{libsodium}.
The \kgen~method of Additive HASE has as parameters the number of ciphertext components and the number of bits per component.
From these, it deterministically derives a set of primes $\{d_1, \ldots, d_t\}$.
The Additive HASE \dec~method computes the discrete logarithms using exhaustive search with a fixed set of precomputed values.

Our compiler converts floating-point plaintexts to a fixed-point representation by an application-defined scaling factor.
It also transforms the calculations to integer values, whereby the scaling factors are considered when appropriate.
The resulting value is transformed back to floating point after decryption.
The compiler then performs the transformation described in Section~\ref{sec:coed} and outputs a \emph{Main Class} containing the program start code, multiple \emph{App Classes} containing the remaining code and \emph{Conversion Data} (\eg labels and comparison parameters)~(4).

% Trusted Module Part I
\paragraph*{Deployment}
Second, the client deploys the app classes at the cloud server and securely loads the generated cryptographic keys and conversion data into the Trusted Module.
We implemented the Trusted Module using an Intel SGX enclave.
SGX is well suited for our implementation, because it provides the following features we require: remote attestation, secure data storage and isolated program execution. 
It is available in every Intel processor beginning at the Skylake generation making it a widely available Trusted Module.
Using the remote attestation feature of SGX the client prepares the enclave (refer to~\cite{Intel_SGX3} for details).
This feature allows to verify the correct creation of an enclave in a remote system and -- in our case -- the correct setup of the crypto library.
Additionally, SGX's remote attestation provides means to establish a secure channel between an external party and an enclave, over which we transfer the sensitive conversion data to the untrusted cloud server.
We emphasize that the cryptographic keys and the conversion data is protected from access by any software except the enclave by SGX's hardware protection.

\subsection{Runtime Phase}
\label{sec:impl-runtime}

\begin{figure*}[t]
\centering
\includegraphics[width=\textwidth]{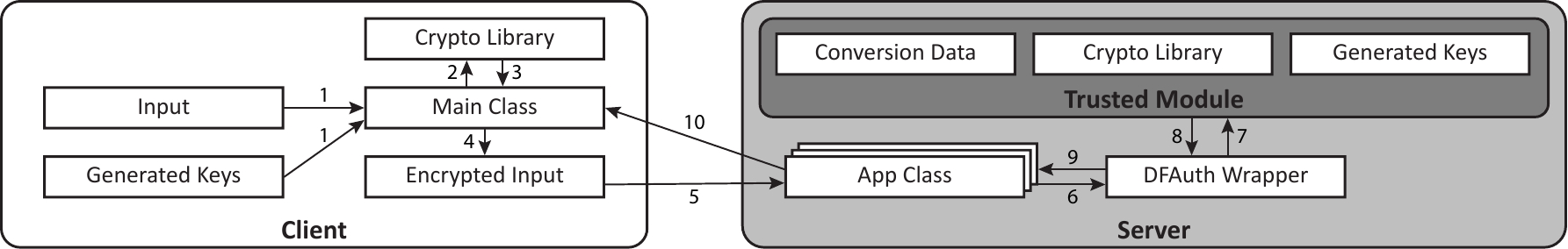}
\caption{
 Application Execution during Runtime Phase. See Section~\ref{sec:impl-runtime} for a full description.
}
\label{fig:impoverview-runtime}
\end{figure*}

To run the program, the client executes the main class which triggers the remote program execution at the cloud server (see Figure~\ref{fig:impoverview-runtime}).
The main class encrypts the {\em Program Input} (for this run of the program) with the generated keys (for the entire setup of the program) using the crypto library (1-4).

The main class passes the {\em Encrypted Input} to the app classes on the cloud server (5).
The app classes operate on encrypted data and do not have any additional protection.
% DFAuth Wrapper
They invoke the \emph{\dfauth Wrapper} for operations on homomorphic ciphertexts and re-encryption or comparison requests to the Trusted Module (6).
The \dfauth wrapper hides the specific homomorphic encryption schemes and Trusted Module implementation details in data flow authentication from the app classes. %, such that it is feasible to run the same bytecode program with data flow authentication on different homomorphic encryption schemes or Trusted Modules.
The wrapper forwards re-encryption and comparison requests to the Trusted Module and passes the answers back to the application (7-9).

Once the app classes have finished their computation, they send an encrypted result (including an authentication label) back to the client (10).
The client verifies the authentication label to the one computed by our compiler.

% Trusted Module Part II
The task of the Trusted Module during runtime is to receive re-encryption and comparison requests, determine whether they are legitimate and answer them if they are.
It bundles cryptographic keys, authentication labels and required parts of the crypto library inside a trusted area shielding it from unauthorized access.
The \dfauth wrapper enables to potentially select different Trusted Modules based on the client's security and feature requirements and their availability at the cloud server.
Besides Intel SGX enclaves, one can implement a Trusted Module using a hypervisor or calling back to the client for sensitive operations.

SGX's secure random number generator provides the random values required during encryption.
A restriction of the current generation of Intel SGX is the limited size of its isolated memory.
It only provides about \SI{96}{\mega\byte} for code and data and thus gives an upper bound for precomputed discrete logarithm values used to speedup Additive HASE.
Exhaustive search is still possible, but our experiments showed that it is severely slower.
The available memory can be used optimally with a careful selection of CRT parameters.
% In our experiments, the calculations with plaintext floats and our encrypted scaled values differs only at \todo{...} decimal places.

Re-encryption and comparison requests have to be implemented inside the Trusted Module.
We display the conversion routines (implemented in an SGX enclave in our case) for conversion to multiplicative homomorphic encryption and comparison in Listings \ref{lst:mul} and \ref{lst:comp}.
The conversion routine to additively homomorphic encryption is similar to the one for multiplicative encryption in Listing \ref{lst:mul} with the roles of the encryption schemes switched.
The comparison of two encrypted values is similar to the comparison of one to a constant in Listing \ref{lst:comp}.
Similar to the call \verb!labelLookup!, which retrieves labels from conversion data stored inside the Trusted Module, \verb!idLookup! and \verb!paramLookup! retrieve identifiers for encryption and parameters for comparison from the conversion data.

\begin{lstlisting}[caption=Conversion to multiplicative HE, label=lst:mul]
convertToMul(x, "x") {
  label = labelLookup("x");
  y = Dec(K, x, label);
  if (y == fail)
    stop;
  id = idLookup("x");
  return Enc(K, y, id);
}
\end{lstlisting}

\begin{lstlisting}[caption=Conversion to comparison, label=lst:comp]
convertToCmp(x, y, "x") {
  label = labelLookup("x");
  x1 = Dec(K, x, label);
  if (x1 == fail)
    stop;
  if (y == null) {
    param = paramLookup("x");
    switch (param.type) {
      case EQ:
        return (x1 == param.const);
      case GT:
        return (x1 > param.const);
      case GTE:
    ...
  } else {
    label = labelLookup("y");
    y1 = Dec(K, y, label);
    if (y1 == fail)
      stop;
    ...
  }
}
\end{lstlisting}

\section{Evaluation}
\label{sec:eval}
In this section, we present the evaluation results collected in two experiments.
In our first experiment, we apply \dfauth to a checkout (shopping cart)
 component of a \emph{Secure Sales Application}, which we developed ourselves.
In our second experiment, we use \dfauth to transform an existing neural
 network program enabling \emph{Secure Neural Networks in the Cloud}.

All experiments were performed on an Intel Core~i7-6700 CPU with
 \SI{64}{\giga\byte} RAM running Windows 10.
For the evaluation, we aimed for a security level equivalent to $80$ bits of
 symmetric encryption security.
We used the \emph{$1536$-bit MODP Group} from RFC3526~\cite{RFC3526} as the
 underlying group in Multiplicative HASE.
The \verb+libsodium+ \cite{libsodium} elliptic curve group used by Additive
 HASE even provides a security level of $128$ bits \cite{Curve25519}.

\subsection{Secure Sales Application}
In this experiment, we consider the checkout component of a secure sales
 application running at an untrusted server.
When a client wants to checkout a shopping cart, the server is tasked with
 summing up the encrypted prices of all items in the cart.
Additionally, discounts need to be applied when the sum exceeds certain thresholds.

In this application we want to protect the discount structure of the client, \ie the thresholds when discounts are applied and the percentage of the discount.
This is important in outsourced shopping applications, because an attacker, \eg a co-residing competitor, could try to infer the discount structure in order to gain an unfair advantage in the market.

\paragraph*{Experimental Setup}
We implemented this checkout component and applied \dfauth to it.
If the sum exceeds the value of \SI[]{500}[\$]{}, we grant a total discount of 10\%.
If the sum exceeds the value of \SI[]{250}[\$]{}, our implementation grants a discount
 of 5\%.

In order to evaluate the performance of the original (\emph{plaintext}) and the
 \dfauth variant of the program, we built shopping carts of sizes $\{1, 10,
 \ldots, 100\}$.
Prices were taken uniformly at random from the interval $[0.01, 1000.00]$.
For each cart a plaintext and an encrypted variant is stored at the untrusted server.

For each of the two program variants, the total runtime of code executing at
 the untrusted server was measured.
The total runtime includes reading the corresponding shopping cart data from
 disk, summing up the prices of all items in the cart and granting discounts
 where applicable.
For the DFAuth variant, we also collected the number of operations performed on
 encrypted data inside and outside of the Trusted Module, as well as the time
 spent invoking and inside the Trusted Module.
Our measurements do not include the setup phase, because it is only a one-time overhead that amortizes over multiple runs.
Furthermore, we do not include network latency in our measurements, since the difference in communication between a program running on plaintext and a program running on encrypted data is very small.

\paragraph*{Evaluation Results}
There are three cases for the control flow in the secure sales application:
\begin{itemize}
 \item Case 1:
  The sum of all item prices neither reaches the first threshold nor the second threshold.
  In this case, the sum of the prices is compared to two different threshold constants.
 \item Case 2:
  The sum of all prices reaches the larger threshold.
  In this case, the sum is compared to one threshold constant and needs to be converted to Multiplicative HASE before being multiplied with the respective discount constant.
 \item Case 3:
  The sum of all prices reaches the lower threshold, but not the larger threshold.
  In this case, the sum is compared to two threshold constants and needs to be converted to Multiplicative HASE before being multiplied with the respective discount constant.
\end{itemize}

\begin{figure}[bt]
\centering
\includegraphics[width=\columnwidth]{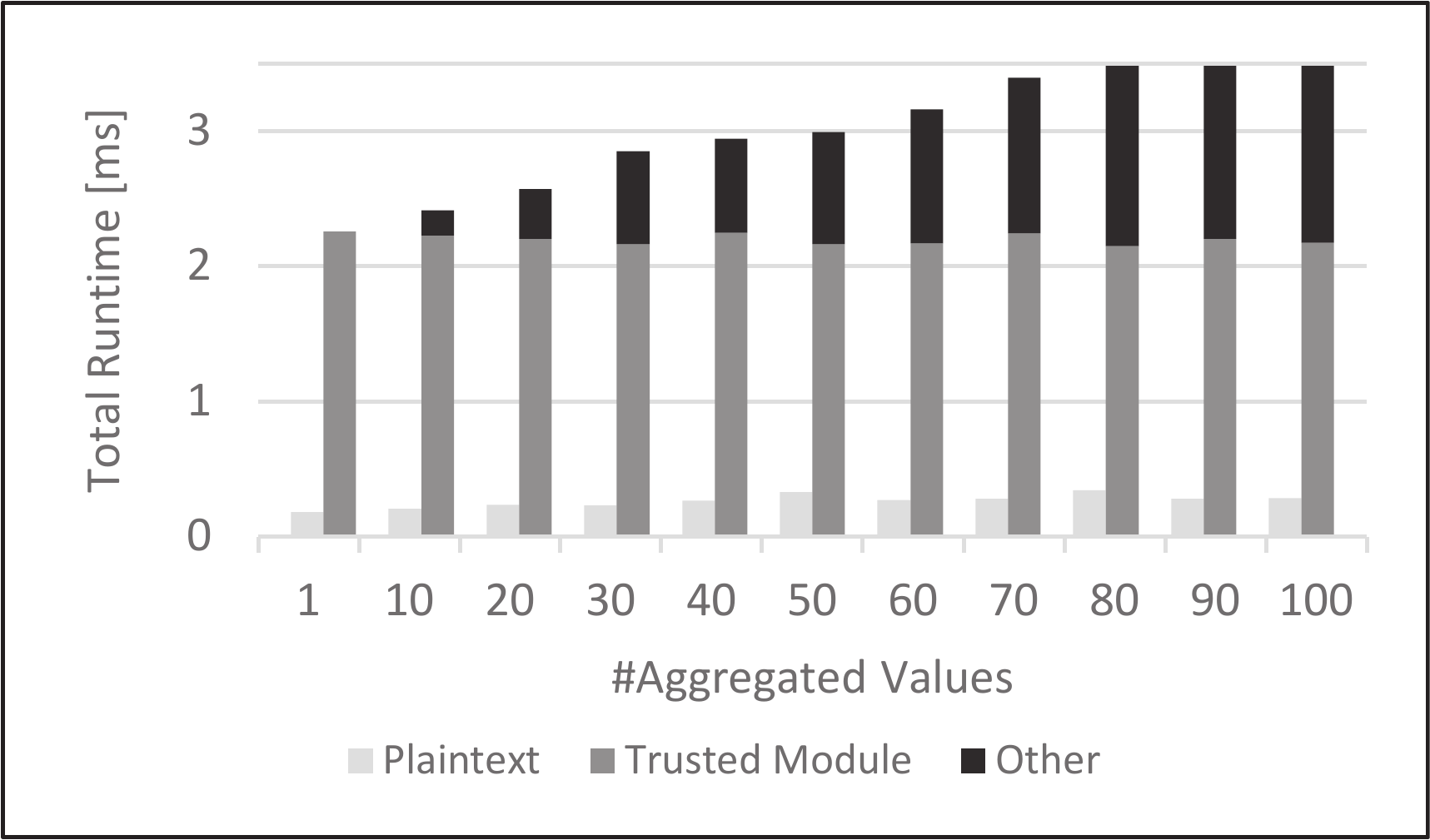}
\caption{
 Average runtime [ms] of the original (left) and \dfauth (right) variants of
  the shopping cart program as a function of the cart size for Case~3.
 We do not show the 95\% confidence interval, because even the largest value is
  only $\pm$~\SI{0.003}{\milli\second}, which would not be visible in the
  graph.
}
\label{fig:barchart-cart-totaltime}
\end{figure}

Figure~\ref{fig:barchart-cart-totaltime} presents the average runtime results
 of $100$ runs of the experiment described above for Case~3.
We can see that even for large-sized carts, which contain 100 items, the
 slowdown caused by the introduction of \dfauth is only about a factor of $3$.
Assuming a roundtrip latency of at least \SI{33}{\milli\second} (which can be considered realistic according to \cite{NetworkAverages}), the slowdown decreases to a factor of $1.3$ at most.
Most importantly, the absolute runtime values are sufficiently low for online computation in a practical deployment.

From Figure~\ref{fig:barchart-cart-totaltime} we can also see that a significant portion of the total runtime is spent inside (or invoking) the Trusted Module.
On the one hand, this shows that a more efficient Trusted Module implementation would significantly decrease the total runtime of the application.
On the other hand, it suggests that we execute more instructions inside the Trusted Module than outside (contradicting our basic idea of a reduced execution inside the Trusted Module).
However, Table~\ref{tab:callComparisonCart}, which reports the number of operations performed on encrypted data inside and outside of the Trusted Module, shows that this is not the case.
Even for a shopping cart containing only $10$ items, there are $9$ to $10$ untrusted HASE operations, but only $1$ to $3$ trusted operations.
%Whether or not $1$ or $2$ comparisons to constants are necessary depends on whether the sum of all item prices is greater than the set thresholds.
%In case the sum is greater than one of the two thresholds, the sum needs to be converted from Additive HASE to Multiplicative HASE.
%Afterwards it can be multiplied with the respective discount using an untrusted homomorphic multiplication.
While the number of trusted operations is independent of the shopping cart size, the number of untrusted HASE operations is approximately linear in the shopping cart size, \ie an even larger shopping cart size would further increase the fraction of untrusted HASE operations.

\begin{table}
\caption{Comparison of the number of Untrusted Operations and Trusted Operations on encrypted data called by the Secure Sales Application experiment for shopping cart size $10$.}
\centering
\ra{1.3}
\begin{tabular}{@{}lrclr@{}}
\toprule
\multicolumn{2}{c}{Untrusted HASE Operations} & \phantom{a} & \multicolumn{2}{c}{Trusted Operations (SGX)} \\
Functionality & \# Ops && Functionality & \# Ops \\
\cmidrule{1-2} \cmidrule{4-5}
Hom. Addition &&& \makecell[lc]{Additive to Multiplicative\\ HASE Conversion} & \\
\makecell[r]{Case 1:} & $9$ && \makecell[r]{Case 1:} & $0$ \\
\makecell[r]{Case 2:} & $9$ && \makecell[r]{Case 2:} & $1$ \\
\makecell[r]{Case 3:} & $9$ && \makecell[r]{Case 3:} & $1$ \\
Hom. Multiplication & && Comparison to Constant & \\
\makecell[r]{Case 1:} & $0$ && \makecell[r]{Case 1:} & $2$ \\
\makecell[r]{Case 2:} & $1$ && \makecell[r]{Case 2:} & $1$ \\
\makecell[r]{Case 3:} & $1$ && \makecell[r]{Case 3:} & $2$ \\
\cmidrule{1-2} \cmidrule{4-5}
Total &&& Total & \\
\makecell[r]{Case 1:} &  $9$ && \makecell[r]{Case 1:} & $2$ \\
\makecell[r]{Case 2:} & $10$ && \makecell[r]{Case 2:} & $2$ \\
\makecell[r]{Case 3:} & $10$ && \makecell[r]{Case 3:} & $3$ \\
\bottomrule
\end{tabular}
\label{tab:callComparisonCart}
\end{table}

\subsection{Secure Neural Networks in the Cloud}
In this experiment, we consider the use case of evaluating neural networks in
 the cloud.
Due to their computational complexity, it is desirable to outsource neural
 network computations to powerful computing machinery located at a cloud
 service provider.
 
In this application we want to protect the neural network model and the instance classified, \ie the weights of the connections and the inputs and outputs of the neurons.
The weights do not change between classifications and often represent intellectual property of the client.
Also the privacy of a user classified in the neural network is at risk, since his classification may be revealed.
Our \dfauth mechanism overcomes these concerns, because it encrypts the weights in the network and the client's input and performs only encrypted calculations.
Note that even the classification does not leak, since the result returned is the output values for each of the classification neurons, \ie a chance of classification $y$, \eg breast cancer in our subsequent example, of $x$\%. 
%\todo{Can not mention breast cancer here because the setup was not introduced yet?}

\paragraph*{Experimental Setup}
We apply our transformation to the BrestCancerSample [sic] neural network provided by Neuroph \cite{Neuroph}, a framework for neural networks.
Given a set of features extracted from an image of a breast tumour, the
 BrestCancerSample neural network predicts whether the tumour is malignant or benign.
As such, the network operates on highly sensitive medical data.

The properties of the neural network (\eg layer and neuron configuration) are encoded programmatically in the Main Class of this network.
This class also reads the data set associated with the network and divides it into a 70\% training set and a 30\% test set.
The training set is used to learn the network, the test set is used
 to evaluate whether the network delivers correct predictions.

We start by applying our \dfauth mechanism to the Main Class of the network and
 the classes of the framework (App Classes).
Result of the transformation is a new Main Class and a set of App Classes
 operating on ciphertexts rather than floating-point double values.
Floating-point numbers are converted to fixed-point numbers by scaling by a factor of $10^6$.
We use the facilities provided by Neuroph to serialize the trained neural
 network weights into a double array and encrypt each weight using HASE.
The encrypted weights and the neural network configuration form the encrypted neural network.
We exploit Neuroph to write the encrypted neural network to disk just like the original one operating on plaintext.

For both -- the plaintext and encrypted neural network -- we test 11 different network evaluation sizes ($\{1, 10, 20, \ldots, 100\}$) and perform $20$ runs each.
For every run, a new random segmentation of training and test data is done and the network is trained again.
Inputs to the neural network are sampled uniformly at random (without replacement) from the test data set.
As in the previous experiment, we measured the total runtime of code executing
 at the untrusted server, the time spent invoking and inside the Trusted Module,
 the number of operations performed on encrypted data inside and outside of the
 Trusted Module.
The total runtime includes reading the network configuration (\ie layers and
 neuron), building the network, loading the weights and executing the network
 evaluation.

\begin{figure}[bt]
\centering
\includegraphics[width=\columnwidth]{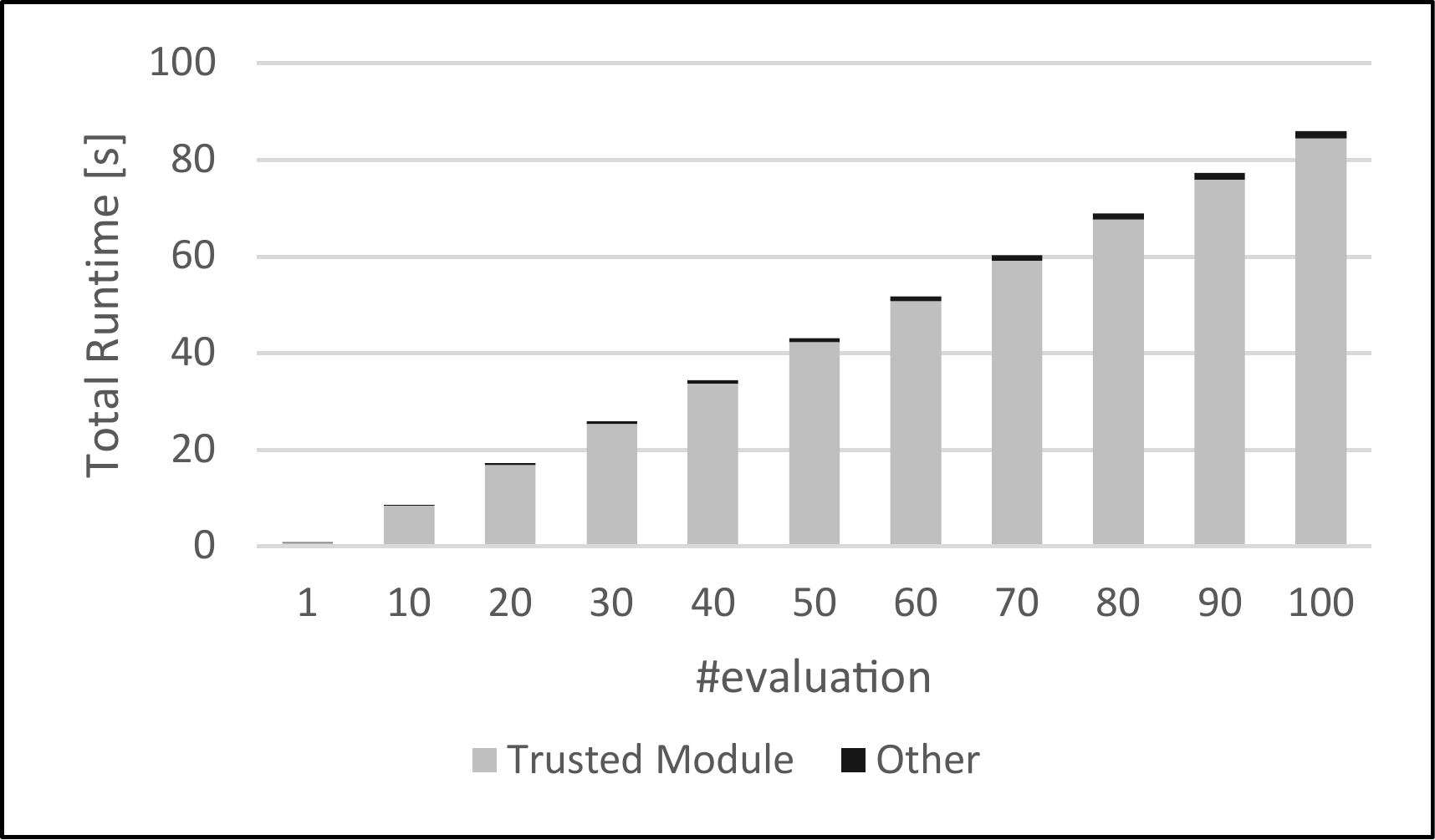}
\caption{
 Average total runtime [s] of the \dfauth variant of the breast cancer neural network experiment as a function of the number of evaluations.
 We do not show the 95\% confidence interval, because even the largest value is only $\pm$~\SI{0.02}{\second}, which would not be visible in the graph.
 For the same reason, we do not include the plaintext measurements.
}
\label{fig:barchart-neuroph-totaltime}
\end{figure}

\paragraph*{Evaluation Results}
We present the evaluation results of the encrypted neural network in Figure~\ref{fig:barchart-neuroph-totaltime}.
The total runtime of one network evaluation is only \SI{0.86}{\second}, whereby \SI{0.84}{\second} (98\%) are spend in the Trusted Module (SGX) and \SI{0.02}{\second} (2\%) on the cloud server, but outside of the Trusted Module.
Even for $100$ evaluations the run completes in \SI{85.96}{\second} on average.
In this case, the processing time in the Trusted Module is \SI{84.39}{\second} (98\%) and \SI{1.56}{\second} (2\%) outside.
The relative runtime of an evaluation (total runtime / number of network evaluations) is constant at about \SI{860}{\milli\second} with a 95\% confidence interval of $\mypm$~\SI{0.42}{\milli\second}.
Clearly, a user can wait online for the output of a neural network evaluation on encrypted data in a practical deployment.
It is important to note that for each run and every input, the prediction of the encrypted network was consistent with the prediction of the plaintext network, \ie we introduce no additional error due to the encrypted computation.
Compared to one plaintext network evaluation, we introduce a slowdown of about $677$.

As in the previous experiment, a large portion of the total runtime is spend inside the Trusted Module.
Table~\ref{tab:callComparisonNeuroph} reports the number of untrusted operations and number of Trusted Module operations.
For a single neural network evaluation, 1096 untrusted operations and 620 trusted operations on encrypted data are performed.
This means that 64\% of all operations on encrypted can be performed directly using homomorphic encryption, without involving the Trusted Module.

\begin{table}
\caption{Comparison of the number of Untrusted Operations and Trusted Operations on encrypted data called for a single evaluation of the neural network.}
\centering
\ra{1.3}
\begin{tabular}{@{}lrclr@{}}
\toprule
\multicolumn{2}{c}{Untrusted HASE Operations} & \phantom{a} & \multicolumn{2}{c}{Trusted Operations (SGX)} \\
Functionality & \# Ops && Functionality & \# Ops \\
\cmidrule{1-2} \cmidrule{4-5}
Hom. Addition       & $548$  && \makecell[lc]{Additive to Multiplicative\\ HASE Conversion} & $36$ \\
Hom. Multiplication & $548$  && \makecell[lc]{Multiplicative to Additive\\ HASE Conversion} & $548$ \\
                    &        && Comparison to Constant                                    & $36$ \\
\cmidrule{1-2} \cmidrule{4-5}
Total             & $1096$   && Total              & $620$ \\
\bottomrule
\end{tabular}
\label{tab:callComparisonNeuroph}
\end{table}

\paragraph*{Comparison to Alternative Solutions}
Recently, implementations of machine learning on encrypted data have been presented for somewhat homomorphic encryption \cite{GilBac16} and Intel's SGX \cite{OhrSch16}.
Compared to the implementation on somewhat homomorphic encryption our approach offers the following advantages:
\begin{itemize}

\item Our approach has a {\em latency} of $0.86$ seconds compared to $570$ seconds for somewhat homomorphic encryption.
The implementation in \cite{GilBac16} exploits the inherent parallelism of somewhat homomorphic encryption in order to achieve a high throughput.
However, when evaluating only one sample on the neural network the latency is large.
Our approach is capable of evaluating only a single sample with low latency as well.

\item Our approach scales to different machine learning techniques with {\em minimal developer effort}.
Whereas the algorithms in \cite{GilBac16} were specifically developed for a type of neural network, our implementation on encrypted data was derived from an existing implementation of neural networks on plaintext data by compilation.
This also implies that the error introduced by \cite{GilBac16} due to computation on integers does not apply in our case.
However, we have not evaluated this aspect of accuracy in comparison to \cite{GilBac16}.

\item Our approach is capable of {\em outsourcing} a neural network evaluation whereas the approach in \cite{GilBac16} is a two-party protocol, \ie the weights of the neural network are known to the server.
Our approach encrypts the weights of the neural network and hence a client can outsource the computation of neural network.
Note that our approach includes the functionality of evaluating on plaintext weights as well and hence offers the larger functionality.

\end{itemize}

Although their runtime overhead is smaller than ours, our approach offers the following advantage compared to the implementation on Intel's SGX \cite{OhrSch16}:
In our approach the code in the SGX enclave is {\em independent of the functionality}, \eg machine learning.
The implementation in \cite{OhrSch16} provides a new, specific algorithm for each additional machine learning function, \ie neural networks, decision trees, etc.
Each implementation has been specifically optimized to avoid side channels on SGX and hopefully scrutinized for software vulnerabilities.
The same development effort has been applied once to our conversion routines and crypto library running in the Trusted Module.
However, when adding a new functionality our approach only requires compiling the source program and not applying the same effort again on the new implementation.

\section{Related Work}
\label{sec:rel}
Our work is related to (homomorphic) authenticated encryption and computation over encrypted data -- including but not limited to homomorphic encryption.

\paragraph*{(Homomorphic) Authenticated Encryption}

Authenticated Encryption (AE) is an encryption mode that provides confidentiality as well as authenticity (unforgeability) and is the recommended security notion for symmetric encryption schemes.
Classically, AE is achieved by composing an encryption scheme providing confidentiality with a message authentication code providing authenticity.
Recent, more efficient encryption schemes provide both properties out of the box, \eg Galois Counter Mode (GCM).

An AE can be obtained by composing an IND-CPA secure encryption scheme with a signature or message authentication code (MAC) \cite{BelNam08}.
Hence, one can obtain a {\em homomorphic} AE by combining a homomorphic encryption scheme with a homomorphic MAC.
However, since the best known homomorphic MACs \cite{GenWic13} are not yet fully homomorphic a different construction is required.
Joo and Yun provide the first fully homomorphic AE \cite{JooYun14}.
However, their decryption algorithm is as complex as the evaluation done on homomorphic ciphertexts undermining the advantages of an encrypted programs, \ie one could do the entire computation in the Trusted Module.
In parallel work, Barbosa et al.~develop labeled homomorphic encryption \cite{BarCat17} which, however, has not been applied to Trusted Modules.
Our goal is to separate between an encrypted program space and small, Trusted Module.

Boneh et al.~\cite{BonFre09} introduced linearly homomorphic signatures and MACs in order to support the efficiency gain by network coding.
However, their signatures were still deterministic, hence not achieving IND-CPA security.
Catalano et al.~\cite{CatMar14} integrated MACs into efficient, linearly homomorphic Paillier encryption \cite{Pai99} and used identifiers in order to support public verifiability, \ie verification without knowledge of the plaintext.
However, their scheme also has linear verification time undermining the advantages of a small Trusted Module.

In our HASE construction we aimed for using identifiers and not plaintext values in order to enable data flow authentication.
Furthermore, we split verification into a pre-computed derivation phase and a verification phase.
Hence, we can achieve constant time verification.

Aggregate MACs \cite{KatLin08} provide support for aggregation of MACs from distinct keys.
However, our current data flow authentication considers one client and secret key.

\paragraph*{Computation over Encrypted Data}

Since fully homomorphic encryption \cite{Gen09} provides rather disappointing performance \cite{GenHal12} researchers have resorted to partially encrypting computations.
MrCrypt \cite{TetLes13} infers feasible encryption schemes using type inference.
In addition to homomorphic encryption, MrCrypt makes use of randomized and deterministic order-preserving encryption.
However, the set of feasible programs is limited and the authors only evaluate it on shallow MapReduce program snippets.
Even, in this case several test cases cannot be executed.
JCrypt \cite{DonMil16} improved the type inference algorithm to a larger set of programs.
However, still no conversions between encryption schemes were performed.

AutoCrypt \cite{ACCCS13} used these conversion, however, realized their security implications.
The authors hence disallowed any conversion from homomorphic encryption to searchable encryption.
This restriction prevents any program from running that modifies its input and then performs a control flow decision.
Such programs include the arithmetic computations we performed in our evaluation.

Next to programs written in imperative programming languages, such as Java, programs in declarative languages, such as SQL are amenable to encrypted computation.
In a declarative language, the programmer does not specify the control flow decisions, but they may be re-ordered (optimized) by the interpreter or compiler.
Hence any resulting data and information is admissible and weaker encryption schemes must be used.
Hacig\"um\"us et al.~used deterministic encryption to implement a large subset of SQL \cite{HacIye02}.
Popa et al.~used also randomized and order-preserving encryption in an adjustable manner \cite{PopRed11}.

Verifiable computation \cite{GenGen10} can be used by a client to check whether a server performed a computation as intended -- even on encrypted data.
However, this does not prevent the attacks by malicious adversaries considered in this paper.
It only proves that the server performed one correct computation, but not that it did not perform any others.

Functional encryption \cite{BonSah11} is a more powerful computation on encrypted data than homomorphic encryption.
It not only can compute any function, but also reveal the result of the computation and not only its ciphertext.
However, generic constructions \cite{GolKal13} are even slower than homomorphic encryption.
Searchable encryption \cite{SonWag00} is a special case of functional encryption for comparisons.
It could be used to implement comparisons in data flow authentication.
However, since the actual comparison time is so insignificant compared to the cryptographic operations, it is more efficient to implement comparison in the Trusted Module as well.

\section{Conclusions}
\label{sec:con}
We introduce the concept of data flow authentication (\dfauth) which prevents an active adversary from deviating from the data flow in an outsourced program.
This in turn allows to safely use re-encryptions between homomorphic and leaking encryption schemes in order to allow a larger class of programs to run on encrypted data where only the executed control flow is leaked to the adversary.
Our implementation of \dfauth uses a novel authenticated, homomorphic encryption scheme and Trusted Modules in an Intel's SGX enclave.
Compared to an implementation solely on fully homomorphic encryption we offer better and actually practical performance and compared to an implementation solely on Intel's SGX we offer a much smaller trusted code base independent of the protected application.
We underpin these results by an implementation of a bytecode-to-bytecode compiler that translates Java programs into Java programs on encrypted data using \dfauth.

%\paragraph{Future Work}
% Notes on CRT Elgamal:
% - The more additions, the 'harder' to decrypt, due to exponent growth.
% - We can 'adjust' the ciphertext size to data type (byte, int, long).
% - We can use program analysis techniques to first find the maximum addition-depth.

%\section{Acknowledgements}
%This work has received funding from the European Union's Seventh Framework
% Programme and Horizon 2020 Research and Innovation Programme under grant
% agreements No.~609611, No.~644579 and No.~644412 of the PRACTICE, ESCUDO-CLOUD
% and TREDISEC projects.

%\begin{acks}
%\end{acks}

\bibliographystyle{IEEEtranS}
\bibliography{\jobname}

% Generated by IEEEtranS.bst, version: 1.14 (2015/08/26)
\begin{thebibliography}{10}
\providecommand{\url}[1]{#1}
\csname url@samestyle\endcsname
\providecommand{\newblock}{\relax}
\providecommand{\bibinfo}[2]{#2}
\providecommand{\BIBentrySTDinterwordspacing}{\spaceskip=0pt\relax}
\providecommand{\BIBentryALTinterwordstretchfactor}{4}
\providecommand{\BIBentryALTinterwordspacing}{\spaceskip=\fontdimen2\font plus
\BIBentryALTinterwordstretchfactor\fontdimen3\font minus
  \fontdimen4\font\relax}
\providecommand{\BIBforeignlanguage}[2]{{%
\expandafter\ifx\csname l@#1\endcsname\relax
\typeout{** WARNING: IEEEtranS.bst: No hyphenation pattern has been}%
\typeout{** loaded for the language `#1'. Using the pattern for}%
\typeout{** the default language instead.}%
\else
\language=\csname l@#1\endcsname
\fi
#2}}
\providecommand{\BIBdecl}{\relax}
\BIBdecl

\bibitem{NetworkAverages}
``At\&t gloabl ip network - network averages,''
  \url{http://ipnetwork.bgtmo.ip.att.net/pws/averages.html}.

\bibitem{MPIR}
``Mpir: Multiple precision integers and rationals,'' \url{http://mpir.org}.

\bibitem{Neuroph}
``Neuroph -- java neural network framework,''
  \url{http://neuroph.sourceforge.net}.

\bibitem{libsodium}
``The sodium crypto library (libsodium),''
  \url{https://download.libsodium.org/doc/}.

\bibitem{alpern1988detecting}
B.~Alpern, M.~N. Wegman, and F.~K. Zadeck, ``Detecting equality of variables in
  programs,'' in \emph{Proceedings of the 15th ACM Symposium on Principles of
  Programming Languages}, ser. POPL, 1988.

\bibitem{Intel_SGX3}
I.~Anati, S.~Gueron, S.~P. Johnson, and V.~R. Scarlata, ``{Innovative
  Technology for {CPU} Based Attestation and Sealing},'' in \emph{Workshop on
  Hardware and Architectural Support for Security and Privacy}, ser. HASP,
  2013.

\bibitem{BarCat17}
M.~Barbosa, D.~Catalano, and D.~Fiore, ``Labeled homomorphic encryption -
  scalable and privacy-preserving processing of outsourced data,'' in
  \emph{Proceedings of the 22nd European Symposium on Research in Computer
  Security}, ser. ESORICS, 2017.

\bibitem{BelNam08}
M.~Bellare and C.~Namprempre, ``Authenticated encryption: Relations among
  notions and analysis of the generic composition paradigm,'' \emph{Journal of
  Cryptology}, vol.~21, no.~4, 2008.

\bibitem{Games}
M.~Bellare and P.~Rogaway, ``Code-based game-playing proofs and the security of
  triple encryption,'' in \emph{Proceedings of the 25th International
  Conference on Advances in Cryptology}, ser. EUROCRYPT, 2006.

\bibitem{Curve25519}
D.~J. Bernstein, ``Curve25519: New diffie-hellman speed records,'' in
  \emph{Public Key Cryptography - {PKC} 2006, 9th International Conference on
  Theory and Practice of Public-Key Cryptography, New York, NY, USA, April
  24-26, 2006, Proceedings}.

\bibitem{BonFre09}
D.~Boneh, D.~Freeman, J.~Katz, and B.~Waters, ``Signing a linear subspace:
  Signature schemes for network coding,'' in \emph{Proceedings of the 12th
  International Workshop on Public Key Cryptography}, ser. PKC, 2009.

\bibitem{BonSah11}
D.~Boneh, A.~Sahai, and B.~Waters, ``Functional encryption: Definitions and
  challenges,'' in \emph{Proceedings of the 8th Theory of Cryptography
  Conference}, ser. TCC, 2011.

\bibitem{206170}
F.~Brasser, U.~M{\"u}ller, A.~Dmitrienko, K.~Kostiainen, S.~Capkun, and A.-R.
  Sadeghi, ``Software grand exposure: {SGX} cache attacks are practical,'' in
  \emph{Proceedings of the 11th {USENIX} Workshop on Offensive Technologies},
  ser. WOOT, 2017.

\bibitem{CatMar14}
D.~Catalano, A.~Marcedone, and O.~Puglisi, ``Authenticating computation on
  groups: New homomorphic primitives and applications,'' in \emph{Proceedings
  of the 20th International Conference on the Advances in Cryptology}, ser.
  ASIACRYPT, 2014.

\bibitem{DonMil16}
Y.~Dong, A.~Milanova, and J.~Dolby, ``Jcrypt: Towards computation over
  encrypted data,'' in \emph{Proceedings of the 13th International Conference
  on Principles and Practices of Programming on the Java Platform}, ser. PPPJ,
  2016.

\bibitem{RFC4634}
D.~{Eastlake 3rd} and T.~Hansen, ``{US Secure Hash Algorithms (SHA and
  HMAC-SHA)},'' RFC 4634 (Informational), Internet Engineering Task Force,
  2006.

\bibitem{Elgamal}
T.~Elgamal, ``A public key cryptosystem and a signature scheme based on
  discrete logarithms,'' \emph{IEEE Transactions on Information Theory},
  vol.~31, no.~4, 1985.

\bibitem{GenGen10}
R.~Gennaro, C.~Gentry, and B.~Parno, ``Non-interactive verifiable computing:
  Outsourcing computation to untrusted workers,'' in \emph{Proceedings of the
  30th International Conference on Advances in Cryptology}, ser. CRYPTO, 2011.

\bibitem{GenWic13}
R.~Gennaro and D.~Wichs, ``Fully homomorphic message authenticators,'' in
  \emph{Proceedings of the 19th International Conference on the Advances in
  Cryptology}, ser. ASIACRYPT, 2013.

\bibitem{Gen09}
C.~Gentry, ``Fully homomorphic encryption using ideal lattices,'' in
  \emph{Proceedings of the Symposium on Theory of Computing}, ser. STOC, 2009.

\bibitem{GenHal12}
C.~Gentry, S.~Halevi, and N.~P. Smart, ``Homomorphic evaluation of the {AES}
  circuit,'' in \emph{Proceedings of the 32nd International Conference on
  Advances in Cryptology}, ser. CRYPTO, 2012.

\bibitem{GilBac16}
R.~Gilad{-}Bachrach, N.~Dowlin, K.~Laine, K.~E. Lauter, M.~Naehrig, and
  J.~Wernsing, ``Cryptonets: Applying neural networks to encrypted data with
  high throughput and accuracy,'' in \emph{Proceedings of the 33rd
  International Conference on Machine Learning}, ser. ICML, 2016.

\bibitem{GolKal13}
S.~Goldwasser, Y.~T. Kalai, R.~A. Popa, V.~Vaikuntanathan, and N.~Zeldovich,
  ``Reusable garbled circuits and succinct functional encryption,'' in
  \emph{Proceedings of the Symposium on Theory of Computing}, ser. STOC, 2013.

\bibitem{HacIye02}
H.~Hacig{\"u}m{\"u}{\c{s}}, B.~Iyer, C.~Li, and S.~Mehrotra, ``Executing sql
  over encrypted data in the database-service-provider model,'' in
  \emph{Proceedings of the ACM International Conference on Management of Data},
  ser. SIGMOD, 2002.

\bibitem{Intel_SGX2}
M.~Hoekstra, R.~Lal, P.~Pappachan, V.~Phegade, and J.~Del~Cuvillo, ``{Using
  Innovative Instructions to Create Trustworthy Software Solutions},'' in
  \emph{Workshop on Hardware and Architectural Support for Security and
  Privacy}, ser. HASP, 2013.

\bibitem{CRT-ELG}
Y.~Hu, W.~Martin, and B.~Sunar, ``Enhanced flexibility for homomorphic
  encryption schemes via crt,'' in \emph{Proceedings (Industrial Track) of the
  10th International Conference on Applied Cryptography and Network Security},
  ser. ACNS, 2012.

\bibitem{JooYun14}
C.~Joo and A.~Yun, ``Homomorphic authenticated encryption secure against
  chosen-ciphertext attack,'' in \emph{Proceedings of the 20th International
  Conference on the Advances in Cryptology}, ser. ASIACRYPT, 2014.

\bibitem{KatLin08}
J.~Katz and Y.~Lindell, ``Aggregate message authentication codes,'' in
  \emph{Proceedings of the Cryptographers' Track of the RSA Conference}, ser.
  CT-RSA, 2008.

\bibitem{Katz:2014:IMC:2700550}
------, \emph{Introduction to Modern Cryptography, Second Edition},
  2nd~ed.\hskip 1em plus 0.5em minus 0.4em\relax Chapman \& Hall/CRC, 2014.

\bibitem{RFC3526}
T.~Kivinen and M.~Kojo, ``More modular exponential (modp) diffie-hellman groups
  for internet key exchange (ike),'' RFC 3526 (Proposed Standard), Internet
  Engineering Task Force, 2003.

\bibitem{SOOT}
P.~Lam, E.~Bodden, O.~Lhotak, and L.~Hendren, ``The soot framework for java
  program analysis: a retrospective,'' in \emph{Cetus Users and Compiler
  Infastructure Workshop}, ser. CETUS, 2011.

\bibitem{MS_SGX_Attack}
J.~Lee, J.~Jang, Y.~Jang, N.~Kwak, Y.~Choi, C.~Choi, T.~Kim, M.~Peinado, and
  B.~B. Kang, ``Hacking in darkness: Return-oriented programming against secure
  enclaves,'' in \emph{Proceedings of the 26th {USENIX} Security Symposium},
  ser. USENIX Security, 2017.

\bibitem{203698}
S.~Lee, M.-W. Shih, P.~Gera, T.~Kim, H.~Kim, and M.~Peinado, ``Inferring
  fine-grained control flow inside {SGX} enclaves with branch shadowing,'' in
  \emph{Proceedings of the 26th {USENIX} Security Symposium}, ser. USENIX
  Security, 2017.

\bibitem{Intel_SGX1}
F.~McKeen, I.~Alexandrovich, A.~Berenzon, C.~V. Rozas, H.~Shafi, V.~Shanbhogue,
  and U.~R. Savagaonkar, ``{Innovative Instructions and Software Model for
  Isolated Execution},'' in \emph{Workshop on Hardware and Architectural
  Support for Security and Privacy}, ser. HASP, 2013.

\bibitem{OhrSch16}
O.~Ohrimenko, F.~Schuster, C.~Fournet, A.~Mehta, S.~Nowozin, K.~Vaswani, and
  M.~Costa, ``Oblivious multi-party machine learning on trusted processors,''
  in \emph{Proceedings of the 25th {USENIX} Security Symposium}, ser. USENIX
  Security, 2016.

\bibitem{Pai99}
P.~Paillier, ``Public-key cryptosystems based on composite degree residuosity
  classes,'' in \emph{Proceedings of the 17th International Conference on
  Theory and Application of Cryptographic Techniques}, ser. EUROCRYPT, 1999.

\bibitem{PopRed11}
R.~A. Popa, C.~M.~S. Redfield, N.~Zeldovich, and H.~Balakrishnan, ``Cryptdb:
  protecting confidentiality with encrypted query processing,'' in
  \emph{Proceedings of the 23rd ACM Symposium on Operating Systems Principles},
  ser. SOSP, 2011.

\bibitem{Schwarz2017}
M.~Schwarz, S.~Weiser, D.~Gruss, C.~Maurice, and S.~Mangard, ``Malware guard
  extension: Using sgx to conceal cache attacks,'' in \emph{Proceedings of the
  14th International Conference on Detection of Intrusions and Malware, and
  Vulnerability Assessment}, ser. DIMVA, 2017.

\bibitem{KeyLength}
N.~Smart, ``Algorithms, key size and parameters report,'' 2014.

\bibitem{SonWag00}
D.~X. Song, D.~Wagner, and A.~Perrig, ``Practical techniques for searches on
  encrypted data,'' in \emph{Proceedings of the 2000 Symposium on Security and
  Privacy}, ser. S\&P, 2000.

\bibitem{TetLes13}
S.~Tetali, M.~Lesani, R.~Majumdar, and T.~Millstein, ``Mrcrypt: Static analysis
  for secure cloud computations,'' in \emph{Proceedings of the ACM
  International Conference on Object Oriented Programming Systems Languages \&
  Applications}, ser. OOPSLA, 2013.

\bibitem{ACCCS13}
S.~Tople, S.~Shinde, Z.~Chen, and P.~Saxena, ``Autocrypt: Enabling homomorphic
  computation on servers to protect sensitive web content,'' in
  \emph{Proceedings of the ACM International Conference on Computer \&
  Communications Security}, ser. CCS, 2013.

\bibitem{WasLohSnelting}
D.~Wasserrab, D.~Lohner, and G.~Snelting, ``On pdg-based noninterference and
  its modular proof,'' in \emph{Proceedings of the 2009 Workshop on Programming
  Languages and Analysis for Security}, ser. PLAS, 2009.

\end{thebibliography}

\appendix
\subsection{Postponed Security Reductions}
\label{app:secred}
\subsubsection{Proof of Theorem~\ref{thm:elg-hase-ind-cpa} (HASE-IND-CPA)}
\label{sec:proof-elg-hase-ind-cpa}
\begin{proof}
Let $\Pi, \mathcal{G}, H$ be as described and let $\adv$ be a PPT adversary.
We use a sequence of games to show that $\adv$'s advantage
 $\advantage{\indcpa}{\adv,\Pi}$ is negligible in $\secpar$.
For Game $n$ we use $S_n$ to denote the event that $b = b'$.
The final game and the encryption oracle used in all games are given in
 Figure~\ref{fig:proof-elg-hase-ind-cpa}.

\textbf{Game 0.}
This is the original experiment from Definition~\ref{def:hase-ind-cpa} except
 that instead of relying on $\Pi$ the challenger performs the exact same
 computations on its own.
Clearly, $\advantage{\indcpa}{\adv,\Pi} = |\prob{S_0} - \frac{1}{2}|$.

% Katz/Lindell p. 83 reduces CPA-security of a symmetric encryption scheme
% construction to the problem of distinguishing a PRF from a random function.
% This proof involves an encryption oracle.

% Katz/Lindell p. 402 reduces CPA-security of the Elgamal public-key encryption
% scheme to the DDH problem. Stebila's 'An introduction to provable security'
% contains a verbose proof on page 9. And so does Victor Shoup's 'Sequence of
% Games' on page 7.
\textbf{Game 1 (Indistinguishability-Based Transition).}
Instead of deriving the label used in the third component of the challenge
 ciphertext using the pseudorandom function $H: \mathcal{K} \times \Isp
 \rightarrow \mathbb{G}$ for some random $k \sample \mathcal{K}$, we make use
 of a random function $f \sample \mathcal{F}$ from the set of functions
 $\mathcal{F} = \{ F : \Isp \rightarrow \mathbb{G} \}$.

We construct a polynomial time algorithm $\bdv$ distinguishing between a
 PRF (for a random key) and a random function using $\adv$ as a black box.
If $\bdv$'s oracle is a pseudorandom function, then the view of $\adv$ is
 distributed as in Game 0 and we have
$
 \prob{S_0} = \prob{\bdv^{\adv,H(k, \cdot)}(\secparam) = 1}
$
for some $k \sample \mathcal{K}$.
If $\bdv$'s oracle is a random function, then the view of $\adv$ is
 distributed as in Game 1 and thus we have
$
 \prob{S_1} = \prob{\bdv^{\adv,f(\cdot)}(\secparam) = 1}
$
for some $f \sample \mathcal{F}$.
Under the assumption that $H$ is a PRF, $|\prob{S_0} - \prob{S_1}|$ is
 negligible.

\textbf{Game 2 (Conceptual Transition).}
Because $f$ is only evaluated on a single input $i$ and $f$ is a random
 function, the result is a random element of $\mathcal{G}$.
Thus, instead of computing $l := f(i)$, we can compute $l := g^s$ for
 a random exponent $s \sample \mathbb{Z}_q$.
Since this is only a conceptual change, we have $\prob{S_1} = \prob{S_2}$.

\textbf{Game 3 (Indistinguishability-Based Transition).}
In the challenge ciphertext we replace $h^r = g^{xr}$ with a random group
 element $g^z$ generated by raising $g$ to the power of a random $z \sample
 \mathbb{Z}_q$.

We construct a polynomial time distinguishing algorithm $\ddv$ solving the DDH
 problem that interpolates between Game 2 and Game 3.
If $\ddv$ receives a real triple $(g^\alpha, g^\beta, g^{\alpha \beta})$ for
 $\alpha, \beta \sample \mathbb{Z}_q$, then $\adv$ operates on a challenge
 ciphertext constructed as in Game 2 and thus we have
$$
 \prob{S_2} = \prob{\ddv^{\adv}(
  \mathbb{G}, q, g, g^\alpha, g^\beta, g^{\alpha \beta}
 ) = 1}
 \text{.}
$$
If $\ddv$ receives a random triple $(g^\alpha, g^\beta, g^{\gamma})$ for
 $\alpha, \beta, \gamma \sample \mathbb{Z}_q$, then $\adv$ operates on a
 challenge ciphertext constructed as in Game 3 and thus we have
$$
 \prob{S_3} = \prob{\ddv^{\adv}(
  \mathbb{G}, q, g, g^\alpha, g^\beta, g^\gamma
 ) = 1}
 \text{.}
$$
In both cases $\ddv$ receives $\Angle{\mathbb{G}, q, g}$ output by
 $\mathcal{G}(\secparam)$.
Under the assumption that the DDH problem is hard relative to $\mathcal{G}$,
 $|\prob{S_2} - \prob{S_3}|$ is negligible.

\textbf{Conclusion.}
In the last game, the first component of the challenge ciphertext is trivially
 independent of the challenge plaintext as well as the challenge identifier.
In the second component, $g^z$ acts like a one-time pad and completely hides
 $m_b$.
Similarly, $l = g^s$ acts like a one-time pad in the third component.
Because the challenge ciphertext does not contain any information about $m_b$
 or $i$, we conclude that $\prob{S_3} = \frac{1}{2}$.
Overall we have that $\advantage{\indcpa}{\adv,\Pi} = \negl$.
\end{proof}

\begin{figure}
\begin{pcvstack}[center]
\procedure[space=auto]
 {$\mathrm{Game_3}_{\adv}^{\indcpa}(\secpar)$}{
 S := \{ \} \\
 \Angle{\mathbb{G}, q, g} \gets \mathcal{G}(\secparam) \\
 a, x, y \sample \mathbb{Z}_q, k \sample \mathcal{K} \\
 h := g^x, j := g^y \\
 ek := \elghaseek \\
 sk := \elghasesk \\
 \Angle{m_0, m_1, i, \state} \gets \adv^{E_{sk}}(\secparam, ek) \\
 \pcif i \in \pi_2(S) \pcthen \\
  \pcreturn 0 \\
 \pcelse \\
  b \sample \bin \\
  S := S \cup \{ (m_b, i) \} \\
  r, \pcbox{s}, \pcbox{z} \sample \mathbb{Z}_q, l := \pcbox{g^s} \\
  c := \Angle{g^r, \pcbox{g^z} \cdot m_b, j^r \cdot {m_b}^a \cdot l} \\
  b' \gets \adv^{E_{sk}}(\secparam, c, \state) \\
  \pcreturn b = b'
}
\pcvspace
\procedure[space=auto]
 {$E_{sk}(m, i)$}{
 \pcparse sk = \elghasesk \\
 \pcif i \in \pi_2(S) \pcthen \\
  \pcreturn \bot \\
 \pcelse \\
  S := S \cup \{ (m, i) \} \\
  r \sample \mathbb{Z}_q, l := H(k, i) \\
  c := \Angle{g^r, h^r \cdot m, j^r \cdot m^a \cdot l} \\
  \pcreturn c
}
\end{pcvstack}
\caption{
 Final security experiment used in HASE-IND-CPA proof.
 Changes compared to the first experiment are highlighted.
}
\label{fig:proof-elg-hase-ind-cpa}
\end{figure}

% KL p. 116: Proof: PRF -> MAC for fixed length MAC
% KL p. 125: Proof: PRF -> CBC-MAC is secure arbitrary length MAC
% KL p. 136: Authenticated Encryption
% KL p. 161: HMAC (no proof). Proof via hash-and-MAC construction on p. 159
\subsubsection{Proof of Theorem \ref{thm:elg-hase-uf-cpa} (HASE-UF-CPA)}
\label{sec:proof-elg-hase-uf-cpa}
\begin{proof}
Let $\Pi, \mathcal{G}, H$ be as described and let $\adv$ be a PPT adversary.
We use a sequence of games to show that $\adv$'s advantage
 $\advantage{\ufcpa}{\adv, \Pi}$ is negligible in $\secpar$.
For Game $n$ we use $S_n$ to denote the event that the adversary wins the game.
The final game is illustrated in Figure~\ref{fig:proof-elg-hase-uf-cpa}.

\textbf{Game 0.}
This is the original experiment from Definition~\ref{def:hase-uf-cpa} except
 that instead of relying on $\Pi$ the challenger performs the exact same
 computations on its own.
Clearly, $\advantage{\ufcpa}{\adv,\Pi} = |\prob{S_0}|$.

\textbf{Game 1 (Conceptual Transition).}
We eliminate the conditional statement by comparing $t$ and $w$ in the return
 statement.

\textbf{Game 2 (Indistinguishability-Based Transition).}
We replace the pseudorandom function $H(k, \cdot)$ with a function $f(\cdot)$
 chosen at random.
Under the assumption that $H$ is a PRF, we have that $|\prob{S_1} - \prob{S_2}|$
 is negligible as in the previous security reduction in
 Theorem~\ref{thm:elg-hase-ind-cpa}.

\textbf{Conclusion.}
We show that $\prob{S_2} = \negl$.
Let X be the event that $\forall i \in I : \exists (m, i) \in S$, \ie all
 identifiers have been used in encryption oracle queries.

In case event X does not happen, the challenger evaluates function $f$ on at
 least one new argument.
By the definition of $f$, the result is a random value in the image of $f$.
This random group element acts as a one-time pad and makes $l$ look random.
Subsequently, $t$ is also random from the point of view of the adversary.
To win the experiment, $\adv$ has to fulfill $t = w$.
Because $t$ is random, $\adv$ can not guess the correct $w$ with
 probability better than $\frac{1}{q}$.
Thus, we have
\begin{align}
 \prob{S_2 \land \neg X} = \frac{1}{q} \cdot \prob{\neg X} \text{ .} \label{eq:P1}
\end{align}
Recall that $q$ is the order of $\mathbb{G}$ (of which $w$ is an element) and
 both are output by the group generation algorithm $\mathcal{G}(\secparam)$.
Also note that $\neg X$ holds when $\adv$ performs no encryption queries at all.

Now consider the case when event X happens and let $\Angle{c, I}$ be the output
 of the adversary.
The set of identifiers $I$ determines a label $l$ and an expected message
 $\tilde{m}$.
Furthermore, let $\tilde{c} = \Angle{\tilde{u}, \tilde{v}, \tilde{w}}$ be
 the ciphertext resulting from the application of $\Pi.\eval$ to ciphertexts
 identified by $I$.
As $\tilde{c}$ is an honestly derived encryption of $\tilde{m}$, the following
 must hold:
\begin{align}
\tilde{m} &= \tilde{u}^{-x} \cdot \tilde{v} \nonumber \\
\tilde{w} &= \tilde{u}^y \cdot \tilde{m}^a \cdot l \nonumber \\
            &= (\tilde{u}^{y-x} \cdot \tilde{v})^a \cdot l \label{eq:X}
\end{align}
Similary, in order for $c = \Angle{u, v, w}$ to be accepted as a forgery
 regarding $I$, it must hold that:
\begin{align}
w &= ({u}^{y-x} \cdot {v})^a \cdot l \label{eq:Y}
\end{align}
for some $m := {u}^{-x} \cdot v \neq \tilde{m}$.
Because $m \neq \tilde{m}$ we know that
 $\tilde{u}^{y-x} \cdot \tilde{v} \neq {u}^{y-x} \cdot {v}$ and
 $\tilde{w} \neq w$.

Combining equations (\ref{eq:X}) and (\ref{eq:Y}) yields
\begin{align}
\dfrac
 {\tilde{w}}
 {w}
&=
\dfrac
 {(\tilde{u}^{y-x} \cdot \tilde{v})^a \cdot l}
 {({u}^{y-x} \cdot {v})^a \cdot l} \nonumber \\
&=
\left(
\dfrac
 {\tilde{u}^{y-x} \cdot \tilde{v}}
 {{u}^{y-x} \cdot {v}}
\right)^a \label{eq:Z}
\end{align}
In order for $c$ to be a forgery with regard to $I$, equation (\ref{eq:Z})
 needs to be satisfied.
But since $a$ is a random element of $\mathbb{Z}_q$, the probability that
 $\adv$ can satisfy (\ref{eq:Z}) is only $\frac{1}{q}$.
Hence,
\begin{align}
 \prob{S_2 \land X} = \frac{1}{q} \cdot \prob{X} \text{ .} \label{eq:P2}
\end{align}

Summing up (\ref{eq:P1}) and (\ref{eq:P2}), we have
$$
\prob{S_2}
 = \prob{S_2 \land \neg X} + \prob{S_2 \land X}
 = \frac{1}{q}
$$
and overall we have that
$
\advantage{\ufcpa}{\adv,\Pi} = \negl
$
\end{proof}

\begin{figure}
\begin{pcvstack}[center]
\procedure[space=auto]
 {$\mathrm{Game_2}_{\adv, \mathcal{G}, H}^{\ufcpa}(\secpar)$}{
 S := \{ \} \\
 \Angle{\mathbb{G}, q, g} \gets \mathcal{G}(\secparam) \\
 a, x, y \sample \mathbb{Z}_q, k \sample \mathcal{K}, \pcbox{f \sample \mathcal{F}} \\
 h := g^x, j := g^y \\
 ek := \elghaseek \\
 sk := \Angle{\mathbb{G}, q, g, a, x, y, h, j, \pcbox{f}} \\
 \Angle{c, I} \gets \adv^{E_{sk}}(\secparam, ek) \\
 l := \prod_{i \in I} \pcbox{f(i)} \\
 \pcparse c = \Angle{u, v, w} \\
 m := {u}^{-x} \cdot v \\
 t := {u}^y \cdot m^a \cdot l \\
 \tilde{m} := \bigoplus_{(m', i) \in S, i \in I} m' \\
 \pcreturn \pcbox{t = w} \land m \neq \tilde{m}
}
\pcvspace
\procedure[space=auto]
 {$E_{sk}(m, i)$}{
 \pcparse sk = \Angle{\mathbb{G}, q, g, a, x, y, h, j, \pcbox{f}} \\
 \pcif i \in \pi_2(S) \pcthen \\
  \pcreturn \bot \\
 \pcelse \\
  S := S \cup \{ (m, i) \} \\
  r \sample \mathbb{Z}_q, l := \pcbox{f(i)} \\
  c := \Angle{g^r, h^r \cdot m, j^r \cdot m^a \cdot l} \\
  \pcreturn c
}
\end{pcvstack}
\caption{
 Final security experiment used in HASE-UF-CPA proof.
 Changes compared to the first experiment are highlighted.
}
\label{fig:proof-elg-hase-uf-cpa}
\end{figure}

\subsubsection{Proof of Lemmata~\ref{lem:ceg-hase-ind-cpa} and \ref{lem:ceg-hase-uf-cpa}}
\begin{proof}[Proof (Sketch)]\renewcommand{\qedsymbol}{}
The security of Construction~\ref{cnst:ceghase}
 (Lemma~\ref{lem:ceg-hase-ind-cpa} and Lemma~\ref{lem:ceg-hase-uf-cpa})
 follows directly from the security of Construction~\ref{cnst:elghase}
 (Theorem~\ref{thm:elg-hase-ind-cpa} and Theorem~\ref{thm:elg-hase-uf-cpa}).
\end{proof}

\end{document}